\documentclass[12pt]{aastex}
\usepackage{natbib}
\usepackage{pdfsync}
\usepackage{amsmath}
\usepackage{tabularx}

\newcommand{\ionhy}{H{\sc ii}}
\newcommand{\neuthy}{H{\sc i}}
\newcommand{\kms}{$\mbox{km~s}^{-1}$}
\newcommand{\msol}{\mbox{M\hbox{$_\odot $}}}

\newcommand{\tf}{\raisebox{-0.6ex}{$\ \stackrel
{\raisebox{-.2ex}{$\textstyle \cdot$}}{\cdot\,\,\cdot}\ $}}

\shorttitle{A 6.7~GHz Methanol Maser Parallax with the LBA}
\shortauthors{Krishnan et al.}

\begin{document}

\title{First Parallax Measurements Towards a 6.7~GHz Methanol Maser with the Australian Long Baseline Array - Distance to G\,339.884$-$1.259.}
\author{V. Krishnan\altaffilmark{1,2},
   S.\ P. Ellingsen\altaffilmark{1},
        M.\ J. Reid\altaffilmark{3},
      A. Brunthaler\altaffilmark{4},
           A. Sanna\altaffilmark{4},
        J. McCallum\altaffilmark{1},
        C. Reynolds\altaffilmark{5},
     H.\ E. Bignall\altaffilmark{5},
    C.\ J. Phillips\altaffilmark{2},
          R. Dodson\altaffilmark{6},
           M. Rioja\altaffilmark{6,7}
     J.\ L. Caswell\altaffilmark{2},
            X. Chen\altaffilmark{8},
      J.\ R. Dawson\altaffilmark{9,2},
        K. Fujisawa\altaffilmark{10},
        S. Goedhart\altaffilmark{11,16},
       J.\ A. Green\altaffilmark{12,2},
       K. Hachisuka\altaffilmark{8,10},
           M. Honma\altaffilmark{13},
          K. Menten\altaffilmark{4},
        Z.\ Q. Shen\altaffilmark{8},
    M.\ A. Voronkov\altaffilmark{2},
       A.\ J. Walsh\altaffilmark{5},
              Y. Xu\altaffilmark{14},
           B. Zhang\altaffilmark{8},
       X.\ W. Zheng\altaffilmark{15}}
\affil{1. School of Physical Sciences, University of Tasmania, Private Bag 37, Hobart, Tasmania 7001, Australia ; Vasaant.Krishnan@utas.edu.au}
\affil{2. CSIRO Astronomy and Space Science, Australia Telescope National Facility, CSIRO, PO Box 76, Epping, NSW 1710, Australia}
\affil{3. Harvard-Smithsonian Center for Astrophysics, Cambridge, Massachusetts 02138, USA}
\affil{4. Max-Plank-Institut f\"{u}r Radioastronomie, Auf dem H\"{u}gel 69, 53121 Bonn, Germany}
\affil{5. International Centre for Radio Astronomy Research, Curtin University, Building 610, 1 Turner Avenue, Bentley WA 6102, Australia}
\affil{6. International Centre for Radio Astronomy Research, The University of Western Australia (M468), 35 Stirling Highway, Crawley WA 6009, Australia}
\affil{7. Observatorio Astron\'{o}mico Nacional (IGN), Alfonso XII, 3 y 5, E-28014 Madrid, Spain}
\affil{8. Shanghai Astronomical Observatory, 80 Nandan Road, Shanghai 200030, China}
\affil{9. Department of Physics and Astronomy and MQ Research Centre in Astronomy, Astrophysics and Astrophotonics, Macquarie University, NSW 2109, Australia}
\affil{10. Department of Physics, Faculty of Science, Yamaguchi University, Yoshida 1677-1, Yamaguchi-city, Yamaguchi 753-8512, Japan}
\affil{11. Hartebeesthoek Radio Astronomical Observatory, PO Box 443, Krugersdorp, 1740, South Africa}
\affil{12. SKA Organisation, Jodrell Bank Observatory, Lower Withington, Macclesfield SK11 9DL, UK}
\affil{13. Mizusawa VLBI Observatory, National Astronomical Observatory of Japan \& Department of Astronomical Science, The Graduate University for Advanced Study, Mitaka 181-8588, Japan}
\affil{14. Purple Mountain Observatory, Chinese Academy of Sciences, Nanjing 210008, China}
\affil{15. Department of Astronomy, Nanjing University, Nanjing 210093, China}
\affil{16. SKA South Africa, 3$^{\mbox{\scriptsize rd}}$ Floor, The Park, Park Road, Pinelands, 7405 South Africa}

\begin{abstract}
We have conducted the first parallax and proper motion measurements of 6.7~GHz methanol maser emission using the Australian Long Baseline Array (LBA). The parallax of G\,339.884$-$1.259 measured from five epochs of observations is 0.48$\pm $0.08 mas, corresponding to a distance of $2.1^{+0.4}_{-0.3}$ kpc, placing it in the Scutum spiral arm. This is consistent (within the combined uncertainty) with the kinematic distance estimate for this source at 2.5$\pm $0.5 kpc using the latest Solar and Galactic rotation parameters. We find from the Lyman continuum photon flux that the embedded core of the young star is of spectral type B1, demonstrating that luminous 6.7~GHz methanol masers can be associated with high-mass stars towards the lower end of the mass range.
\end{abstract}

\keywords{astrometry -- Galaxy:kinematics and dynamics -- masers -- stars:formation}

\section{Introduction}
To properly understand the Milky Way's scale and shape as well as the physical properties of the objects within it, including their mass, luminosity, ages and orbits, we need to be able to accurately measure distances to Galactic sources. Unfortunately, this fundamental attribute is one of the most difficult measurements to make. Common techniques include the use of a Galactic rotation model, or ``standard candles'' where the distances to objects with known luminosities can be determined from their observed brightness. These indirect methods of measurement can result in large uncertainties unless an accurate, reliable and robust model is available. For example \citet{Xu+06} found that the kinematic distance was factor of 2 greater than the parallax distances to W3(OH) due to the peculiar velocity of the source. Uncertainties and errors in distance determination naturally propagate into the estimation of other important properties, such as luminosity ($L \propto D^{2}$) and mass ($L \propto M^{a}$, where $3 < a < 4$ for main sequence stars), and hence there is a need to constrain any errors in distance as much as possible. One of the most direct methods to determine distances beyond our solar system is through the use of trigonometric parallax.

A decade ago, \citet{Honma+04} used the VLBI Exploration of Radio Astrometry (VERA) array to conduct high precision astrometry on the water maser pair W49N/OH43.8$-$0.1 and achieved an accuracy of 0.2 mas. This success demonstrated the feasibility of performing maser astrometry on Galactic scales and subsequently several groups obtained parallax results to masers and their associated star formation regions \citep{Hachisuka+06,Xu+06,Honma+07}.

Following this, the BeSSeL project was launched as a comprehensive Northern hemisphere survey to measure accurate distances to high-mass star formation regions (HMSFR) and associated \ionhy \/ regions of our Milky Way galaxy using trigonometric parallax \citep{Reid+09a,Moscadelli+09,Xu+09,Zhang+09,Brunthaler+09,Reid+09b,Sanna+09,Brunthaler+11}. This is being undertaken using the National Radio Astronomy Observatory's (NRAO) Very Long Baseline Array (VLBA) to measure position changes of methanol and water masers in the Galactic disk, with respect to distant background quasars. Phase referencing of one of the quasar and maser emission data to the other \citep{Alef+88,Beasley+95,Reid+14b}, combined with careful calibration of atmospheric effects, allows the change in the relative separation between the masers and quasars to be accurately measured. Repeated observations timed to maximise the measured amplitude of the parallax signature can measure the parallax to an accuracy of up to 10 $\mu$arcsec and simultaneously determine the source proper motions to $\sim $1~$\mu $arcsec~year$^{-1}$ \citep{Reid+14b}. Thus far the combination of all astrometric VLBI observations (including the European VLBI Network (EVN))\citep[e.g.][]{Reid+14b,Chibueze+14,Imai+13,Honma+12,Sakai+12,Rygl+10,Rygl+08} has yielded more than 100 parallax and proper motion measurements for star forming regions across large portions of the Milky Way visible from the  Northern hemisphere. Having determined the position at a reference epoch, parallax and proper motion of the masers, the complete three-dimensional location and velocity vectors of these sources relative to the Sun can be found. \citet{Reid+14a} used these measurements to fit a rotating disk model of the Milky Way and from this data were able to refine the best-fit Galactic parameters finding the circular rotation speed of the Sun, $\Theta _0 = 240 \pm 8 $~\kms \/ and distance to the Galactic centre $R_0 = 8.34 \pm 0.16$ kpc.

To date, parallax distances to masers have only been measured using Northern hemisphere VLBI arrays, and therefore, the sources for which accurate distances have been measured are heavily concentrated towards objects in the first and second quadrants of the Galaxy. The sources with trigonometric parallax measurements compiled in \citet{Reid+14a} predominantly lie within the Galactic longitude range of $0^{\circ} < l < 240^{\circ}$.  There are two sources within the fourth quadrant of the Galaxy, with the most southerly of these at a declination of $-39^{\circ }$. Sources at southerly declinations achieve only low elevations for Northern hemisphere antennas, and for these, the measurement error in the zenith tropospheric delay estimate produces significant degradation in the astrometric accuracy, and corresponding parallax measurement \citep{Honma+08}. Therefore, the updated estimates for fundamental Galactic parameters such as $\Theta _0$ and $R_0$ are based on data which are restricted to only a little over half the total Galactic longitude range. In order to ensure that models of Galactic structure and rotation are reliable, they must be derived from a more uniformly sampled distribution including sources in the third and fourth quadrants.

Distances to southern maser sources and their associated \ionhy\ regions have exclusively been determined via indirect methods, such as through kinematic distance estimates \citep[e.g.][]{Caswell+75,Caswell+10}, with the ambiguity being resolved with \neuthy \/ absorption against \ionhy \/ continuum \citep{Jones+12,Jones+13}, \neuthy \/ Self-absorption (\neuthy SA) \citep{Green+11} and radio recombination lines \citep{Sewilo+04}. However, there are issues which can produce significant errors in these techniques including, potentially large peculiar motions, broad spectral line profiles, and the as yet unknown relationship between cold \neuthy \/ distribution against the maser associated component in \neuthy SA emission.

In 2008 a project was initiated to precisely measure the positions of maser sources relative to extragalactic quasars using the
Australian Long Baseline Array (LBA). The aim of this ongoing project is to determine the parallax distances to 30 prominent southern HMSFRs. Current maser parallax measurements made with the VLBA and VERA have primarily been towards either 12.2~GHz methanol masers or water masers. However, not all LBA antennas have receivers capable of observing at 12.2~GHz, and many of the LBA antennas have significantly poorer sensitivity at 22~GHz than they do at frequencies less than 10~GHz. Hence the 6.7~GHz class II methanol maser transition was considered to provide the best targets for parallax observation with the LBA. They have strong, stable and compact emission over the timescales required for parallax measurements, and are well sampled from the Southern hemisphere, with close to 1000 sources being documented in the Methanol Multibeam (MMB) survey \citep[][a sensitive, unbiased search for 6.7~GHz methanol masers in the southern Galactic plane]{Caswell+10,Green+10,Caswell+11,Green+12a}. Here we present the first trigonometric parallax measurements of a southern 6.7~GHz methanol maser.

\section{Overview of G\,339.884$-$1.259}
\begin{figure}
  \centering
  \includegraphics[width=\linewidth]{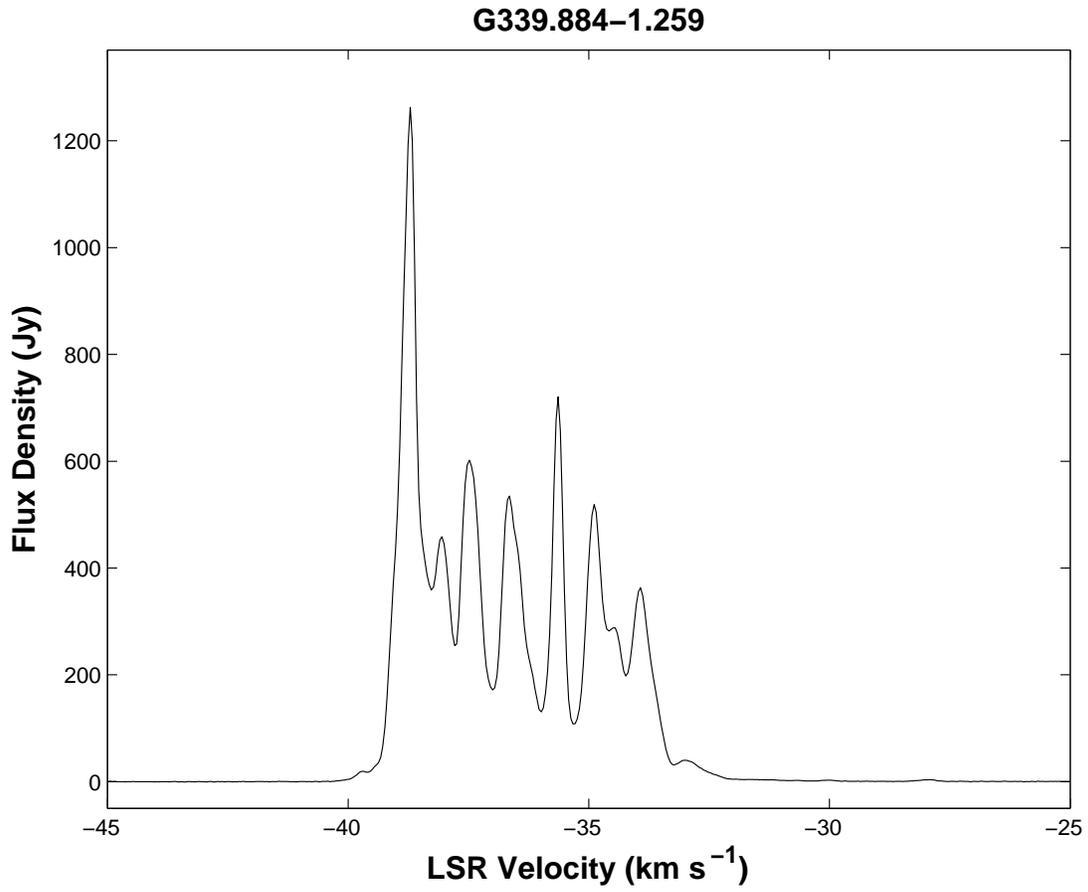}
\caption{The autocorrelation spectrum of G\,339.884$-$1.259 using all antennas except HartRAO from 2013 June.}
  \label{fig:spec}
\end{figure}

G\,339.884$-$1.259 is one of the strongest (1520 Jy at $-$38.7~\kms \/ \citep{Caswell+11}) 6.7~GHz methanol masers and has been well studied in a range of maser and thermal molecular transitions \citep[e.g.][]{Norris+93,deBuizer+02,Ellingsen+04,Ellingsen+11a} and in continuum emission at a range of wavelengths \citep[e.g.][]{Ellingsen+96,Walsh+98,Ellingsen+05}. The autocorrelation spectrum of the 6.7~GHz methanol maser emission of this source from the 2013 June epoch of observation is shown in Figure~\ref{fig:spec}. Its peak flux density and spectral profile has been relatively stable over the last 20 years, and interferometric observations of the masers indicate several compact features \citep{Norris+93,Ellingsen+96,Caswell+11}, making it a suitable candidate for phase referenced astrometry. Previous estimates of the distance to G\,339.884$-$1.259 have utilised the kinematic distance method \citep[e.g.][]{Caswell+83,Green+11} which yields a kinematic heliocentric distance of between 2.5 to 3.0 kpc (depending on the assumed Galactic parameters). The coordinates used for G\,339.884$-$1.259 in the current observations are given in Section~\ref{sec:updatedCoords} with updated coordinates presented in Table~\ref{tab:coords}.

\begin{table*}
\centering
\caption{Coordinates of observed sources. The separation and position angle columns describe the offset between the respective quasar and G\,339.884$-$1.259 in the sky. The reported positions of G\,339.884$-$1.259 and J\,1654$-$4812 have been revised (see Section~\ref{sec:updatedCoords}) based on the 2013 March epoch. We failed to detect J\,1648$-$4826, J\,1644$-$4559, J\,1648$-$4521 and J\,1649$-$4536 in our observations. The upper limits for detection is 5 times the image RMS (from a box of size 1.5$\times $1.5 arcsec).}
\begin{tabularx}{\textwidth}{lccccccr}
\hline
     \bf Source  & \bf Correlated  &\bf Separation  & \bf Position & \bf RA      & \bf Dec  \\
                 & \bf flux        &                & \bf angle & & \\
                 & \bf (mJy)       &\bf ($^{\circ}$) & \bf ($^{\circ}$) & \bf (h m s) & \bf ($^{\circ}$~\arcmin ~\arcsec)   \\
\hline
\hline
 Maser:               &        &      &        &              & & \\
 G\,339.884$-$1.259  & $-$    &  $-$ & $-$    &16 52 04.6776 & $-$46 08 34.404 \\
Detected quasars:              &      &        &              & & & \\
 J\,1706$-$4600      &  131.1 & 2.48 &\,~88.1 &17 06 22.0503 & $-$46 00 17.824 \\
 J\,1654$-$4812      & ~~~3.8 & 2.11 &  169.9 &16 54 18.2448 & $-$48 13 03.756 \\
Non-detected quasars: &        &      &        &              & & \\
 J\,1648$-$4826      & $<$0.5 & 2.36 &  193.3 &16 48 47.9200 & $-$48 26 18.800 \\
 J\,1644$-$4559      & $<$0.5 & 1.27 &  276.5 &16 44 49.2856 & $-$45 59 09.646 \\
 J\,1648$-$4521      & $<$0.9 & 1.02 &  319.1 &16 48 14.2110 & $-$45 21 38.090 \\
 J\,1649$-$4536      & $<$0.4 & 0.72 &  317.1 &16 49 14.7810 & $-$45 36 31.190 \\
\hline
\end{tabularx}
  \label{tab:coords}
\end{table*}

Methanol masers in G\,339.884$-$1.259 were first observed by \citet{Norris+87} at 12.2~GHz and high resolution synthesis images were made of the emission at 6.7, and 12.2~GHz by \citet{Norris+93}. Interestingly, the maser emission was found to have a linear spatial distribution and had a corresponding monotonic velocity gradient. In light of this, G\,339.884$-$1.259 became a prime candidate for astronomers to model the conditions for high-mass star formation \citep{Ellingsen+96,Norris+98,Phillips+98,Walsh+98,deBuizer+02,Dodson+08}. \citet{Ellingsen+96} proposed that the masers are located within a cicrumstellar disk, and \citet{Norris+98} showed that the emission fit a model of a Keplerian disk around an OB type star. However, these claims were disputed by \citet{deBuizer+02} who demonstrated that what was initially thought to be a single circumstellar disk could be resolved into three mid-infrared sources near the location of the methanol masers. \citet{Dodson+08} attempted to test the hypothesis of a circumstellar disk in G\,339.884$-$1.259 by making polarization measurements of the magnetic fields associated with this source. From their images, they report a primarily disordered field accompanying much of the emission, and proposed that this matches the expectations for the masers being associated with an outflow-related shock. One small region of methanol maser emission does show a magnetic field direction perpendicular to the elongation of the maser emission, suggestive of a disk \citep{Dodson+08}. However, the mass of the enclosed object (assuming Keplarian rotation) is of magnitude of only 0.03 $M_{\odot }$ (assuming a distance of 3 kpc to G\,339.884$-$1.259).

A large number of additional class~II methanol maser transitions have been detected towards G\,339.884$-$1.259. Observations of multiple maser transitions are required to place constrains on the environmental conditions in the region surrounding the young high-mass star where the masers arise. This is because any model will need to account for the specific conditions required for the observed transition \citep{Cragg+92,Sobolev+97}. \citet{Norris+93} made high-resolution maps of the 6.7 and 12.2~GHz maser spots in G\,339.884$-$1.259 and found little evidence for spatial coincidence with the spots at different frequencies. \citet{Caswell+00} made observations of 107.0 and 156.6~GHz methanol masers in G\,339.884$-$1.259 and found the emission peak at these frequencies coincided with the 6.7~GHz maser site to within 5$^{\prime \prime }$. These are extremely rare transitions, with only 22 and 4 detections respectively from a pool of 80 sources. \citet{Ellingsen+04} discovered emission from the 19.9~GHz transition in G\,339.884$-$1.259, and in \citet{Krishnan+13} it was shown that there was little evidence for spatial or velocity coincidence between the masers at 6.7 and 19.9~GHz, with only 2~maser components identified at 19.9~GHz as opposed to $\sim $10~components in 6.7 and 12.2 GHz observations with similar angular resolution \citep{Norris+93}.

Ultracompact (UC) \ionhy \/ regions are bubbles of ionized gas associated with newly formed massive stars, and the first detection of such a region in G\,339.884$-$1.259 was by \citet{Ellingsen+96}. The emission was measured at 8.5~GHz and had a peak brightness of 6.1~mJy~beam$^{-1}$. The peak of the methanol emission at $-$38.7~\kms \/ was found to be offset from the continuum peak by 0.6$^{\prime \prime }$, with the methanol masers lying in a line approximately across the centre of the continuum emission and in an orientation which is perpendicular to the direction of extension of the UC\ionhy \/ region. \citet{deBuizer+02} interpret the radio continuum emission is being due to an ionized outflow along the axis of extension.

\section{Observations}
\label{sec:Obs}
The Long Baseline Array (LBA) is a heterogeneous VLBI array with either 5 or 6 antennas available for the observations reported here. In the period spanning 2012 March to 2014 March, a total of six epochs of observations were undertaken (LBA experiment code v255, epochs q to v inclusive) towards the southern 6.7~GHz methanol maser G\,339.884$-$1.259 (see Table~\ref{tab:antennas}). The LBA antennas available for one or more epochs were the Australia Telescope Compact Array (ATCA), Ceduna 30m, Hartebeesthoek 26m, Hobart 26m, Mopra 22m and Parkes 64m antennas. The ATCA is itself a connected 6-element interferometer, which was operated in tied-array mode, with the outputs from either four or five 22~metre antennas phased and
combined.

\begin{table*}
\centering
\caption{Stations which participated in the observations of G\,339.884$-$1.259 and J\,1706$-$4600.}
\begin{tabular}{rccll}\hline
  \bf D.O.Y & \bf Start UT  & \bf Code & \bf Epoch &  \bf Participating stations \\
\hline
\hline
 67 &  04:00 &  v255q  &  2012 March    &  ATCA, Ceduna, Hobart, Mopra, Parkes \\
 77 &  04:00 &  v255r  &  2013 March    &  ATCA, Ceduna, HartRAO, Hobart, Parkes \\
168 &  02:30 &  v255s  &  2013 June     &  ATCA, Ceduna, HartRAO, Hobart, Mopra, Parkes \\
226 &  18:00 &  v255t  &  2013 August   &  ATCA, Ceduna, HartRAO, Hobart, Mopra, Parkes \\
323 &  12:00 &  v255u  &  2013 November &  ATCA, Ceduna, Hobart, Mopra, Parkes \\
 60 &  22:00 &  v255v  &  2014 March    &  ATCA, Ceduna, HartRAO, Hobart, Mopra \\
\hline
\end{tabular}
  \label{tab:antennas}
\end{table*}

The observations for each epoch typically lasted for between 18 to 24 hours, of which approximately one-third of the time was utilised for observation of G\,339.884$-$1.259 (and associated calibration observations). The remaining time was used for observations of other sources, the results of which will be reported in future publications. The setup consisted of two different frequency configurations to maximize on the sensitivity requirements of the different modes. The first configuration was to record continuum observations for calibration of the tropospheric component of the delay, and the second for phase referenced maser observations.

The first configuration utilized as broad a frequency range as possible. The heterogeneous nature of the array meant that with different receiver front ends, it was necessary to compromise the frequency setup so that it could be adopted at all antennas.  The LBA Data Acquisition System (DAS) can record the observed signals onto two independent IF bands. The optimal frequency configuration which is able to accommodate these restrictions is to have 4 $\times$ 16 MHz bands, the first pair (LBA DAS IF 1 with RR polarization) with lower-band edges at 6300 and 6316 MHz and the second pair (LBA DAS IF 2 with LL polarization) with lower-band edges at 6642 and 6658 MHz. Brief observations of approximately 12 to 18 quasars ($\sim $2 minutes per source) with point-like structure at high resolution were undertaken for a broad azimuth range (generally at low elevation). These quasars were selected from the International Celestial Reference Frame (ICRF) Second Realization catalogue \citep{Ma+09}. These ``ICRF observations'' were used to determine the troposphere path length contributions to the delay ($\tau$), as well as to model the clock drift rate at the observatories (Section~\ref{sec:icrfDataRed} and Appendix~\ref{sec:icrfObs}). The ICRF observations were grouped into 45 minute blocks with an interval of between 3 to 6 hours separating consecutive blocks.

The second frequency configuration was for phase referencing between the maser G\,339.884$-$1.259 and a nearby background quasar for parallax determination. The LBA DAS system was set to record dual circular polarisation for 2 $\times $ 16 MHz bands with lower-band edges at 6642 and 6658 MHz. The phase referencing technique was employed by alternating scans for 2~minutes on the target maser with scans lasting 2 minutes on a nearby quasar. In order to achieve accurate phase referenced astrometry, suitable background quasars had to fulfill the criteria that they have little or no extended structure, their coordinates be known to an uncertainty of $<$1~mas and that they are in close angular proximity to the associated maser of $\sim $1$^{\circ}$ \citep{Reid+09a}. The primary databases which were used in our search for quasars were the AT20G \citep{Hancock+11}, Astrogeo \citep{Petrov+11} and ATPMN \citep{McConnell+12} catalogues. A list of potential background quasars which were observed in conjunction with G\,339.884$-$1.259 is given in Table~\ref{tab:coords}. During the data reduction process, the phase information from the maser emission in a single channel (where the emission is strong and compact) is transferred to the nearby quasar (see Section~\ref{sec:dataRed}). In doing this, it is assumed that the state of the array has remained constant in the time interval between consecutive maser scans, so that the quasar phase can be interpolated with small error contributions \citep{Fomalont+13}.

The data were correlated with the DiFX correlator at Curtin University \citep{Deller+11}. As the maser emission covers only a small fraction of the 32 MHz of recorded bandwidth, it is not necessary to correlate the entire band. Hence, for the maser data a 2~MHz zoom-band was correlated with 2048 channels (1 MHz over 1024 channels for the 2014 March epoch), giving a spectral resolution of 0.977~kHz corresponding to a velocity separation of 0.055~\kms \/.  For the observations of the background quasar the entire observed bandwidth was correlated. In the 2012 March and 2013 March epochs, 256 spectral channels were used per 16~MHz bandwidth, corresponding to a resolution of 62.5~kHz. In the remaining epochs, 32 spectral channels were used per 16~MHz bandwidth, corresponding to a resolution of 500~kHz. The ICRF data were correlated using the same spectral resolution as the background quasar observations in each epoch.

\section{Calibration}
We used the Astronomical Image Processing System (AIPS) \citep{Greisen+03} for data processing, employing the same data reduction steps across all epochs and based on the procedure described in \citet{Reid+09a}.

\subsection{ICRF Data}
\label{sec:icrfDataRed}
Prior to using observations of the ICRF catalogue quasars to determine the tropospheric and clock delay parameters for each antenna, it was necessary to remove the estimated ionospheric delay (determined from global models based on GPS total electron content (TEC) observations \citep{Walker+99}), the Earth Orientation Parameters (EOPs) and parallactic angle effects. We then performed delay calibration, using a strong ICRF source with well known position ($<$1 mas) \citep{Ma+09} (see Table~\ref{tab:icrfDelCals} for details). A least squares fit based on the approach in \citet{Reid+09a} was then employed to determine the zenith atmospheric delay and the results (detailed in Appendix~\ref{sec:icrfObs}) were applied to the phase referenced observations in Section~\ref{sec:dataRed}.

\begin{table*}
  \centering
  \caption{Sources in each epoch which were used for ICRF mode delay calibration.}
\begin{tabular}{lcccrl}\hline
 \bf Epoch & \bf Source  & \bf RA         & \bf Dec                & \bf Scan      \\
           & \bf name    & \bf (h m s)    & \bf ($^{\circ}$~\arcmin ~\arcsec )   & \bf (min)          \\
\hline
\hline
2012 March    & 1349$-$439 &   13 52 56.54 & $-$44 12 40.388 & 1:00 \\
2013 March    & 0537$-$441 &   05 38 50.36 & $-$44 05 08.939 & 1:00 \\
2013 June     & 1302$-$102 &   13 05 33.02 & $-$10 33 19.428 & 2:00 \\
2013 August   & 0013$-$005 &   00 16 11.09 & $-$00 15 12.445 & 2:00 \\
2013 November & 1929$+$226 &   19 31 24.92 & \,~~22 43 31.259& 1:15 \\
2014 March    & 0537$-$441 &   05 38 50.36 & $-$44 05 08.939 & 4:00 \\
\hline
\end{tabular}
   \label{tab:icrfDelCals}
\end{table*}

\subsection{Phase Referenced Data}
\label{sec:dataRed}
We removed modeled residuals attributed to the ionospheric delay, EOPs and parallactic angle before applying the troposphere and clock drift rate corrections which were determined from the ICRF observations. The AIPS task CVEL was then used to re-position the maser spectrum in the bandpass to account for Doppler shifts due to antenna positions and Earth's rotation specific to each epoch. Figure~\ref{fig:spec} shows that the G\,339.884$-$1.259, 6.7~GHz maser spectral peak, has a local standard of rest velocity~($v_{\mbox{\scriptsize lsr}}$) of $-$38.8~\kms \/ as reported by \citet{Caswell+11}.

ACCOR was then used to correct the amplitude of the data for imperfect sampler statistics at recording, and also for incorrect amplitude scaling at the correlator which was a problem in some versions of the DiFX software used for correlation. Following this, we extracted a single autocorrelation scan of the maser spectrum from the most sensitive single antenna (Parkes, with the exception of the final epoch) and used ACFIT to scale the spectra at all observing stations to this template, based on the nominal system equivalent flux density (SEFD) of the antennas\footnote[1]{\scriptsize www.atnf.csiro.au/vlbi/documentation/vlbi\_antennas/index.html}. The resultant amplitude gains as a function of time were then applied to the maser and quasar datasets. The calibrated peak flux intensity for the feature at $-$35.6~\kms \/ in G\,339.884$-$1.259 (the maser component used for phase referencing) and J\,1706$-$4600 (the phase referenced quasar used for parallax measurement) for each epoch is presented in Table~\ref{tab:parFitStats}. The RMS noise in the quasar image was determined from an area of size $1.5 \times 1.5$ arcsec and from a region of the image where there was no emission. The RMS noise for the maser image was determined from an area of the same dimensions and from a spectral channel where there was no emission.

\begin{table*}
\centering
\caption{Measured fluxes and differential fitted positions between the $-$35.6 \kms\/ feature in G\,339.884$-$1.259 and J\,1706$-$4600 across all epochs after phase referencing. The formal errors in each coordinate are determined from the output of IMFIT.}
\begin{tabular}{lcccccccccc}\hline
& & & & & \multicolumn{2}{c}{\bf J\,1706$-$4600} & \multicolumn{2}{c}{\bf G\,339.884$-$1.259} \\
\bf Epoch      & \bf x Offset& \bf Error & \bf y Offset & \bf Error & \bf Flux & \bf RMS & \bf Flux & \bf RMS\\
               & \multicolumn{2}{c}{\bf (mas)} & \multicolumn{2}{c}{\bf (mas)} & \multicolumn{2}{c}{\bf (mJy)} & \multicolumn{2}{c}{\bf (Jy)}  \\
\hline
\hline
2012 March    &  4.177  &  0.056   &  2.998  &  0.098   &  30.29  & 0.55 & 413.07 & 0.03 \\
2013 March    &  2.667  &  0.026   &  1.085  &  0.023   &  28.98  & 0.36 & 418.17 & 0.02 \\
2013 June     &  1.654  &  0.003   &  0.464  &  0.002   & 114.84  & 0.15 & 427.49 & 0.02 \\
2013 August   &  1.031  &  0.021   &  0.208  &  0.017   &  61.06  & 0.65 & 462.02 & 0.02 \\
2013 November &  0.727  &  0.016   &  0.006  &  0.012   &  62.25  & 0.38 & 268.95 & 0.02 \\
\hline
\end{tabular}
  \label{tab:parFitStats}
\end{table*}

\begin{figure}
  \centering
     \includegraphics[width=\textwidth]{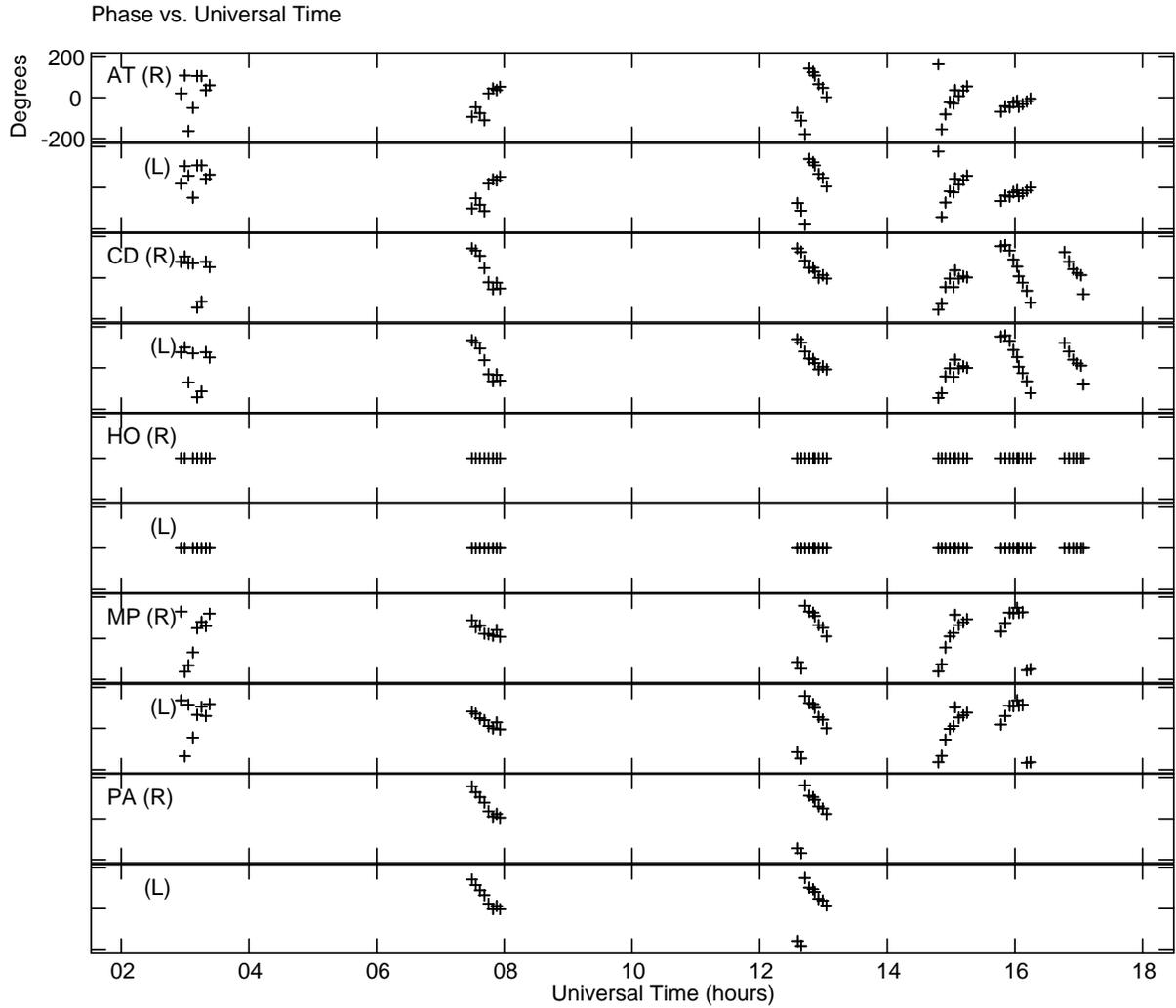}
\caption{Phase solutions in each polarization from ATCA (AT), Ceduna (CD), Hobart (HO), Mopra (MP) and Parkes (PA) in the 2013 August session after fringe fitting (with reference to Hobart 26m) on the $-$35.6 \kms\/ feature in G\,339.884$-$1.259.}
  \label{fig:phases}
\end{figure}

We used a single scan (1 to 2 minutes) on J\,1706$-$4600 for delay calibration to correct the initial residual clock and instrumental error from correlation. However the quasar dataset was not correlated with the same spectral resolution as G\,339.884$-$1.259 (see Section~\ref{sec:Obs}) and there was a need to account for this by modifying the solutions copied to the zoom-band maser data. This crucial step was required in order to prevent the introduction of spurious R-L polarization phase differences into the dataset, and the efficacy of this procedure (see Appendix~\ref{sec:delayCal}) is demonstrated by the similarity in the phase solutions between the different polarizations in Figure~\ref{fig:phases}. Similar phase transfer issues between datasets with differing frequency properties have previously been resolved using comparable methods \citep{Rioja+08,Dodson+14}.

\section{Astrometry, Parallax and Proper Motion}
\label{sec:paraRes}
\begin{figure}
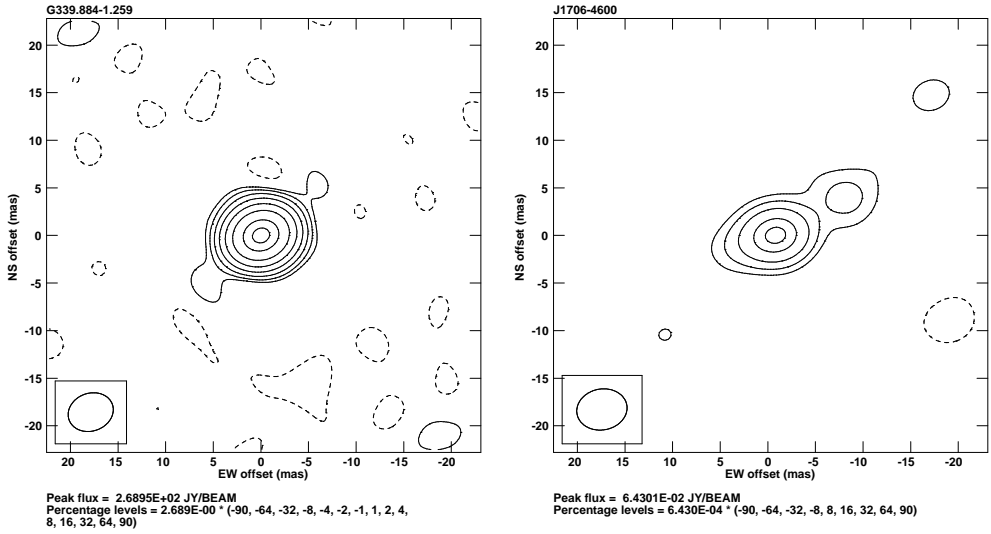

  \centering
       \includegraphics[width=0.4\textwidth]{v255u_G339.eps}
       \includegraphics[width=0.4\textwidth]{v255u_J1706.eps}
\caption{Emission from the phase reference channel corresponding to $v_{\mbox{\scriptsize lsr}} = -$35.6~\kms \/ in G\,339.884$-$1.259 (left) that was strong and showed compact structure. J\,1706$-$4600 (right) showed consistent centroid structure dominated by a single peak throughout all epochs.}
  \label{fig:v255uQSO}
\end{figure}

Figure~\ref{fig:spec} shows that there are multiple strong ($>$100~Jy) 6.7~GHz methanol maser components in G\,339.884$-$1.259 which offer potential spectral channels for astrometry. We examined the cross correlation spectra to find the spectral features which showed the smallest relative flux density variations across all baselines for the duration of the observations, taking this to be an indication that it was an unresolved point source, which would enable an accurate position determination. We found the spectral feature at $-$35.6~\kms \/ (Figure~\ref{fig:v255uQSO}) to be clearly the best choice for G\,339.884$-$1.259. All the other peaks exhibited substantial variability across individual baselines suggesting that emission is not point-like on milliarcsecond scales, or that there is a blend of emission from different locations. We then fringe fit on the maser spectral channel associated with this feature before transferring the phase solutions to J\,1706$-$4600. Figure~\ref{fig:phases} shows a typical plot of phase versus time from the 2013 August observations.

After transferring the phase corrections, we averaged all channels in the quasar dataset and then imaged the emission using a Gaussian beam of $5.9 \times 4.2$~mas (average from all epochs) (Figure~\ref{fig:v255uQSO}). We report detections for J\,1706$-$4600 and J\,1654$-$4812 on VLBI baselines, and only the first of these was suitable for astrometry. J\,1706$-$4600 was observed in all epochs and appears to be dominated by a single component with no jets. J\,1654$-$4812 was observed in all epochs except 2012 March and showed variable source structure which made it unsuitable for parallax determination. J\,1706$-$4600 has a positional accuracy of 2.10~mas \citep{Petrov+11} and shows deviation from point-like structure at levels $<$10\% of the peak flux density (Figure~\ref{fig:v255uQSO}). J\,1654$-$4812 had an estimated positional accuracy of 0.4\arcsec \/ \citep{McConnell+12} and from our phase referenced images, we are able to present updated coordinates for J\,1654$-$4812 to an uncertainty of 2.1~mas in Table~\ref{tab:coords}. We located the centroid position of the quasar by fitting a 2D Gaussian to the deconvolved J\,1706$-$4600 emission. The offset of the emission peak from the centre of the image field was recorded for all epochs and we present these data in Table~\ref{tab:parFitStats}.

\begin{figure}
   \includegraphics[width=\textwidth]{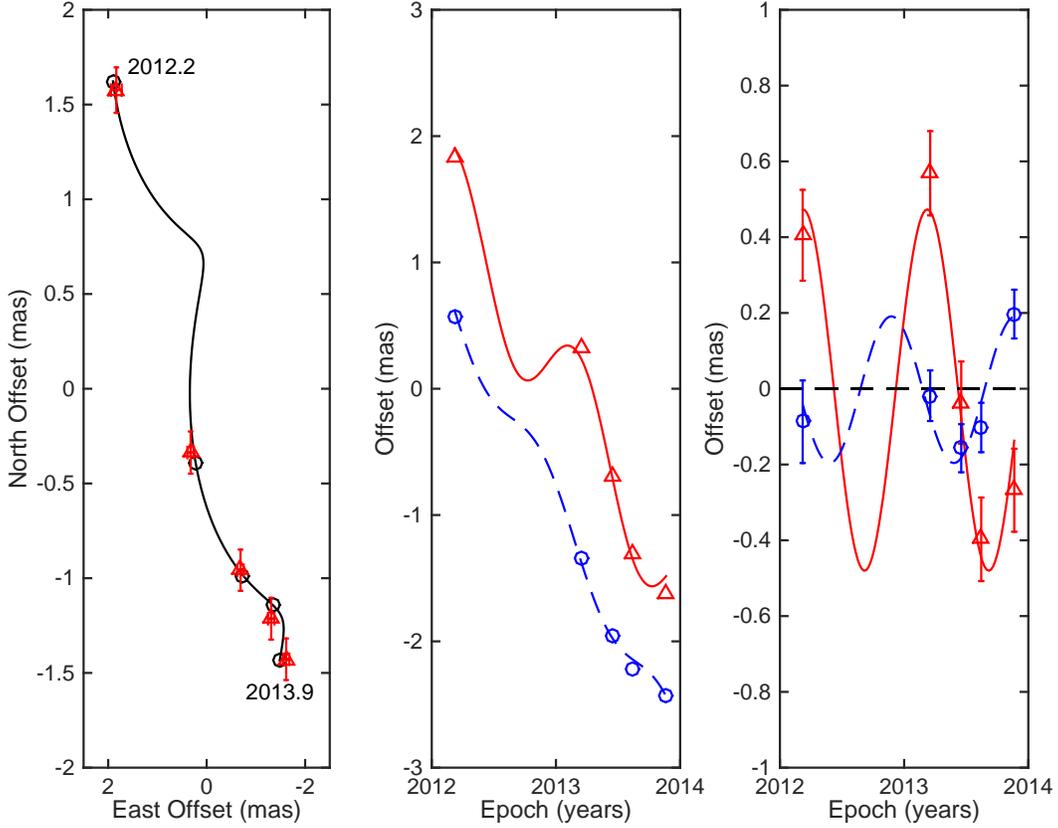}
\caption{Parallax and proper motion of the $-35.6$ \kms\/ reference feature of G\,339.884$-$1.259. The expected positions from the fits are indicated with triangular and circular markers. Left panel: the sky positions with the first and last epochs labeled. Middle panel: East-West (triangles) and  North-South (circles) motion of the position offsets and best combined parallax and proper motions fits versus time. The models are offset along the y-axis for clarity. Right panel: the parallax signature with the best fit proper motions removed.}
  \label{fig:G339_Parallax}
\end{figure}

The change in the position of the $-$35.6~\kms \/ feature in G\,339.884$-$1.259 with respect to J\,1706$-$4600 was modelled independently in right ascension and declination, and included the ellipticity of Earth's orbit. We allowed for systematic sources of uncertainty in right ascension and declination in the parallax model and added these in quadrature to the formal errors in Table~\ref{tab:parFitStats}. A $\chi ^2 _\nu $ (per degree of freedom) for the East-West and  North-South residuals was determined, and we iteratively adjusted the estimated error floors until $\chi ^2 _\nu \approx 1$. The parallax was measured to be 0.48$\pm $0.08 mas corresponding to a distance of $2.1^{+0.4}_{-0.3}$ kpc to G\,339.884$-$1.259. The proper motion was found to be $\mu_{x} = -$1.6$\pm $0.1~mas~y$^{-1}$ and $\mu_{y} = -$1.9$\pm $0.1~mas~y$^{-1}$ (Figure~\ref{fig:G339_Parallax} and Table~\ref{tab:parRes}). In order to constrain errors in the measured proper motion, we made image cubes of the maser emission for all epochs, and analyzed the changes in the distribution from 2012 March to 2014 November. We found that internal motions in G\,339.884$-$1.259 can be up to $\sim $0.4~mas~y$^{-1}$ in right ascension and declination. These motions dominate over the formal errors in $(\mu_x , \mu_y )$ when added in quadrature, and we report the measured proper motion with errors as $\mu_{x} = -$1.6$\pm $0.4~mas~y$^{-1}$ and $\mu_{y} = -$1.9$\pm $0.4~mas~y$^{-1}$. This measured uncertainty corresponds to internal motions of $\sim $2~\kms \/ in the maser emission and is consistent with proper motion estimates of 6.7~GHz methanol masers in HMSFRs \citep[e.g.][]{Goddi+11,Moscadelli+14,Sugiyama+14}. A more detailed analysis of the internal motions of the 6.7~GHz emission in G\,339.884$-$1.259 is beyond the scope of the current text and will be presented in future publications.

\begin{table}
\centering
\caption{Parallax distance and proper motion of G\,{339.884-1.259}.}
\begin{tabular}{cccc}\hline
\bf Maser feature  & \bf Distance & \bf $\mu _x$ & \bf $\mu _y$ \\
\bf (\kms )        & \bf (kpc)    & \bf (mas~y$^{-1}$) & \bf (mas~y$^{-1}$) \\
\hline
\hline
$-$35.6         &  $2.1^{+0.4}_{-0.3}$  &  $-$1.6$\pm $0.1 & $-$1.9$\pm $0.1\\
\hline
\end{tabular}
\label{tab:parRes}
\end{table}

We excluded data from the 2014 March observations in our parallax and proper motion analysis as there were significant technical difficulties during that session which prevented us from obtaining accurate measurements of the maser-quasar separation for that epoch.

The LBA has previously been used to measure distances to a number of pulsars at 1.6~GHz using trigonometric parallax \citep{Dodson+03,Deller+09a,Deller+09b}. The uncertainty in our LBA parallax measurement to G\,339.884$-$1.259 is estimated to be 80 $\mu$arcsec and is equivalent to the errors of the best LBA southern pulsar parallax measurement \citep{Deller+09b}. The uncertainty from our observations is a factor of 2~poorer than the parallax of 6.7~GHz methanol masers in ON~1 ($0.389 \pm 0.045$ mas) measured by \citet{Rygl+10} using the EVN. The LBA (when operating without HartRAO; see Appendix~\ref{sec:icrfObs}) and the EVN configuration in \citet{Rygl+10} both have a maximum East-West baseline separation of $\sim $2000 km, giving them similar resolution capabilities for parallax determination \citep{Reid+09a}. \citet{Rygl+10} reduced random errors in their measurements by modelling parallaxes determined from the averaged positions of several maser spots. However, given the strong and compact nature of the maser spot at $-$35.6~\kms , we are not able to reduce our astrometric errors using this method as these will be dominated by systemics from the uncompensated atmosphere. Additionally, the the EVN observations of ON1 utilise background quasars with separation angles of 1.71$^\circ $ and 0.73$^\circ $, and in comparison J\,1706$-$4600 is separated by 2.48$^\circ $ from G\,339.884$-$1.259 (Table~\ref{tab:coords}). This would have adversely affected the interpolated phase transfer solutions (described in Section \ref{sec:Obs}) and contributed to the larger uncertainty estimate presented here (see Appendix~\ref{sec:icrfObs}). In considering these factors, we assess that with better atmospheric calibration and smaller separation between the maser and quasar, it will be possible to attain parallax accuracies of $\sim $20 $\mu $arcsec using the LBA.

\subsection{Kinematic distance to G\,339.884$-$1.259}
\label{sec:kinDist}
The kinematic distance to sources in the Galactic disk can be determined from a model which describes the rotational speed of the disk at the Sun ($\Theta _0$), distance of the Sun from the Galactic centre ($R_0$), and the measured $v_{\mbox{\scriptsize lsr}}$. \citet{Green+11} report a kinematic distance to G\,339.884$-$1.259 of $2.6 \pm 0.4$ kpc using $\Theta _0 = 246$~\kms , $R_0 = 8.4$~kpc \citep{Reid+09b} and $v_{\mbox{\scriptsize lsr}} = -34.3$~\kms \/ (the mid-point of the 6.7~GHz methanol maser emission range Figure~\ref{fig:spec}). Previous studies of the high-mass star formation region associated with G\,339.884$-$1.259 \citep[e.g.][]{Ellingsen+96,deBuizer+02,Dodson+08} have used a kinematic distance of $\sim $3~kpc to this source. This value was determined using earlier models with Galactic rotation speeds of $\Theta _0 \simeq 220$~\kms \/ (see Section~\ref{sec:pcis} for further comments).

Using updated Galactic parameters of $\Theta _0 = 240$~\kms \/, $R_0 = 8.34$~kpc and solar motion parameters of $U_\odot $ = 10.70~\kms \/ (towards the Galactic centre), $V_\odot $ = 15.60~\kms \/ (clockwise and in the direction of Galactic rotation as viewed from the North Galactic Pole) and $W_\odot $ = 8.90~\kms \/ (towards the North Galactic Pole) \citep{Reid+14a}, we report the kinematic distance to the associated CS(2-1) cloud with $v_{\mbox{\scriptsize lsr}} = -31.6$~\kms \/ \citep{Bronfman+96} to be 2.5$\pm $0.5~kpc. This result is comparable to the parallax distance in Section~\ref{sec:paraRes} but with a large estimated error. Given an uncertainty of $\sim $8~\kms \/ in $\Theta _0$ \citep{Reid+14a}, the estimated error in the quoted kinematic distance of 2.5~kpc is doubled to $\sim $1~kpc.

\subsection{Association of G\,339.884$-$1.259 High-Mass Star Formation Region with Scutum Arm}
Our ability to precisely determine the structure of the Milky Way is hampered by our location in the midst of its spiral arms. \citet{Westerhout+57,Cohen+80,Dame+01,Jones+13}; and others have used surveys of \neuthy \/ and CO molecular clouds to identify the Galaxy's spiral arms from the ensuing longitude-velocity ($\ell - V$) diagrams. This method of spiral arm modelling has been used by \citet{Xu+13,Zhang+13,Choi+14,Reid+14a,Sanna+14,Sato+14,Wu+14} to assign HMSFRs to spiral arms, by associating them with molecular clouds in our Galaxy.

\begin{figure}
  \centering
     \includegraphics[width=\linewidth ]{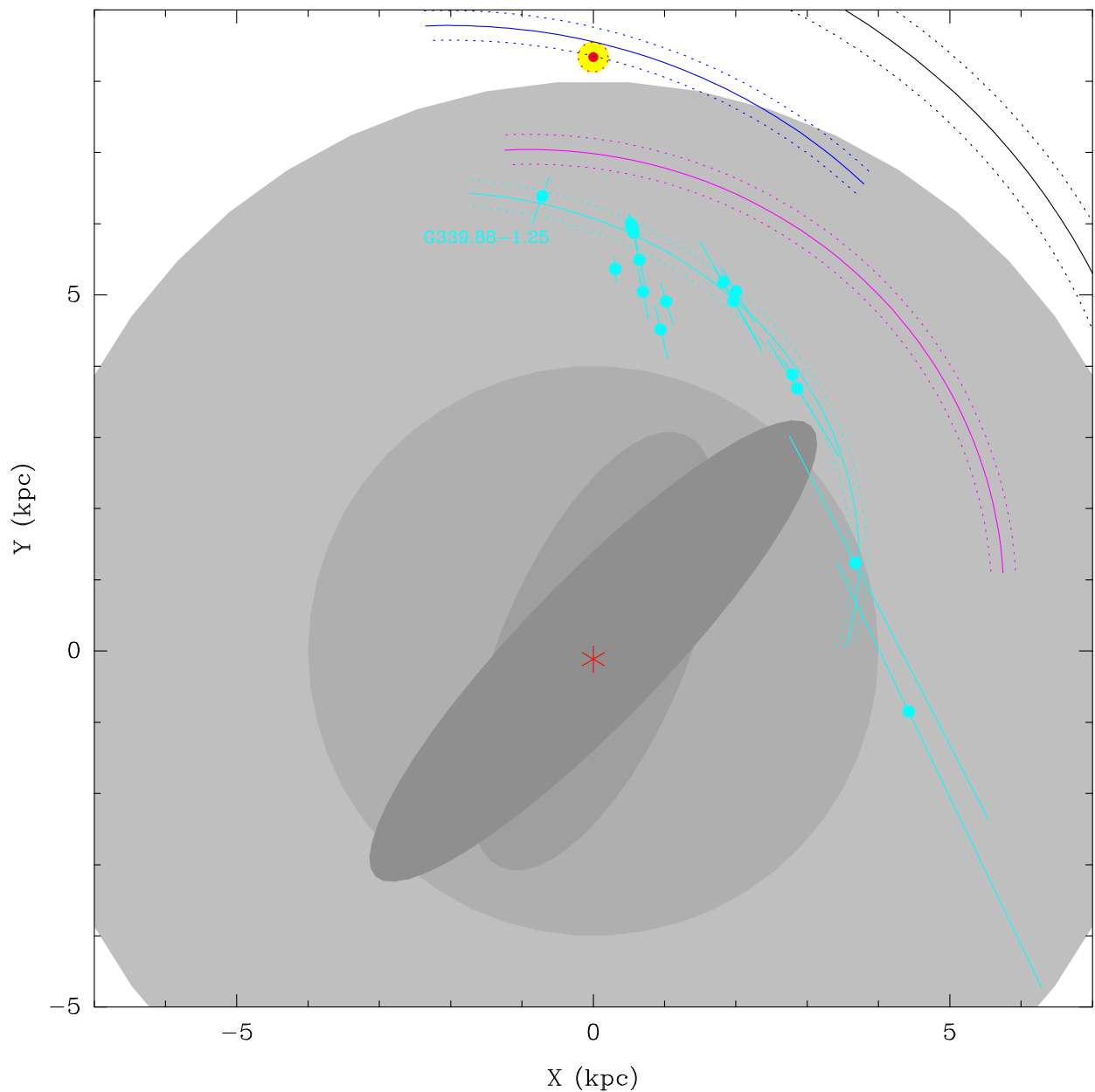}
\caption{A face-on view of the Milky Way galaxy showing G\,339.884$-$1.259 in relation to other HMSFRs in the Scutum arm \citep{Sato+14}. Sections of arcs of the Perseus (top right), Local (top centre with Solar position) and Sagittarius arms are also shown. The background circular disks are scaled to approximate the Galactic bar region ($\sim $4 kpc) and the solar circle ($\sim $8 kpc) (see \citet{Reid+14a}).}
  \label{fig:scutum}
\end{figure}

Based on the $v_{\mbox{\scriptsize lsr}}$ of the CO spectrum in the region (T. Dame 2014, private communication) and the parallax distance in Table~\ref{tab:parRes}, we suggest that G\,339.884$-$1.259 is located in the near edge of the Scutum spiral arm as shown in Figure~\ref{fig:scutum}. \citet{Sato+14} modeled the Scutum arm based on measurements of 16 HMSFRs, and we show the position of G\,339.884$-$1.259 with relation to these in Figure~\ref{fig:scutum}. It can be seen that at the longitude of G\,339.884$-$1.259, the Sagittarius and Scutum arms are in close proximity, and extrapolating the best information currently available suggests that these two arms may merge at lower longitudes.

Using a log-periodic spiral model, \citet{Sato+14} measured a pitch angle of $\psi = $ 19.8$^\circ \pm $3.1$^\circ $ for the Scutum arm. We have now included G\,339.884$-$1.259 into this model, using a source which extends the Galactocentric azimuth by about 10$^\circ $, giving an updated value of $\psi = $ 19.2$^\circ \pm $4.1$^\circ $.

\subsection{Updated coordinates for G\,339.884$-$1.259 and J\,1654$-$4812}
\label{sec:updatedCoords}
During the observations we used $\alpha =$ 16$^{h}$52$^{m}$04.6700$^{s}$, $\delta = -$46$^{\circ}$08\arcmin 34.200\arcsec \/ as the coordinates for G\,339.884$-$1.259 \citep{Caswell+11}. This is the position of the maser spectral peak at $-$38.7~\kms \/ (Figure~\ref{fig:spec}). When applying phase corrections from the $-$35.6~\kms \/ feature to J\,1706$-$4600, we can assume that any offset of the quasar image from the centre of the field is due to the offset of the phase referenced position with respect to the true maser position. We iteratively corrected the maser coordinates in the AIPS source table using CLCOR until there was no further improvement in the quasar position from the image centre in the 2013 March epoch of observations. It is important to use accurate maser coordinates in the data reduction process, to minimize errors in determining the quasar position during phase referencing \citep{Reid+09a}. The position corrections from the March 2013 epoch of observations were then applied to G\,339.884$-$1.259 for all epochs, and we present the updated coordinates corresponding to the $-$38.7~\kms \/ feature in Table~\ref{tab:coords}. We find that there is a difference of 0.219\arcsec \/ between our measured position and that reported by \citet{Caswell+11}. The formal errors in the fitted position for J\,1706$-$4600 were found to be $<$0.1~mas (Table~\ref{tab:parFitStats}). As this is an order of magnitude smaller than the known positional error of this source (see Section~\ref{sec:Obs}), the resultant uncertainty in the updated maser position remains as 2.1~mas when error contributions are added in quadrature.

Observations of J\,1654$-$4812 were made using the coordinates $\alpha =$ 16$^h$54$^m$18.24$^s$, $\delta =-$48$^\circ $13\arcmin 03.7\arcsec \/ from \citet{McConnell+12}. The updated position of this source when measured relative to the corrected G\,339.884$-$1.259 position is also presented in Table~\ref{tab:coords} to an accuracy of 2.1 mas. There is a separation difference of 0.074\arcsec \/ between the original and updated positions.

\section{Properties of associated high-mass star formation region}
\subsection{Peculiar motion}
The proper motions $\mu _x$ = $-$1.6$\pm $0.4~mas~y$^{-1}$, $\mu _y$ =$-$1.9$\pm $0.4~mas~y$^{-1}$ and of $v_{\mbox{\scriptsize lsr}}$ = $-$31.6~\kms \/ of the CS(2-1) cloud associated with this source \citep{Bronfman+96}, make it possible to determine the full 3-dimensional motion of G\,339.884$-$1.259 with respect to the Galactic centre. The dynamical model of the Galaxy we use assumes a flat rotation curve of the disk with a speed of $\Theta _0 = 240$ \kms. This is a reasonable assumption based on the recent analysis in \citet{Reid+14a}. The distance of the Sun to the Galactic centre is taken to be $R_0 = 8.34$ kpc, and assumed to have peculiar motion components $U_\odot $ = 10.70~\kms , $V_\odot $ = 15.60~\kms \/ and $W_\odot $ = 8.90~\kms \/ \citep{Reid+14a}. When using this model, we find the peculiar motion for G\,339.884$-$1.259 to be U = $-$4.0$\pm $5.9~\kms , $V$ = 6.47$\pm $4.6~\kms \/ and $W$ = 10.0$\pm $1.2~\kms , in a reference frame that is rotating with the Galaxy. Hence all components of peculiar velocity are consistent (within estimated uncertainty) with the bulk of HMSFRs in the Milky Way. In \citet{Reid+14a} there is a good fit to the model of spiral arm motions when a RMS of about 5 - 7~\kms \/ is assumed for each velocity component of HMSFR. This is reasonable for Virial motions of stars in giant molecular clouds and so there may not be much evidence for large ($>$10~\kms ) streaming motions in general.

\subsection{Physical constraints on the ionizing star}
\label{sec:pcis}
High-mass star formation is still not well understood, and there is a vibrant debate regarding the processes which result in their formation \citep[e.g.][]{McKee+03,Bonnell+98,Garay+99}. Accurate distances to star formation regions help to put tight constraints on the physical environments from which young high-mass stars are born. This includes fundamental attributes such as size, enclosed mass and luminosity. Therefore, any accurate model describing the star formation process must also be governed by the limits of these physical constraints.

Previous groups have determined the properties of
G\,339.884$-$1.259 based on kinematic distance (see Section~\ref{sec:kinDist}). Using values from \citet{Ellingsen+96} and the equations from \citet{Panagia+78}, we present updated estimates for the electron density~($n_e$), mass of ionized hydrogen~($M_{\mbox{\scriptsize \ionhy }}$), excitation parameter~($U$) and the Lyman continuum photon flux~($N_L$) in Table~\ref{tab:starType}. Based on $\log N_L = $ 45.6~s$^{-1}$  in Table~\ref{tab:starType}, we find that continuum emission is consistent for a core object to be a B1 type star \citep{Panagia+73}. This classification implies that G\,339.884$-$1.259 is relatively small for a young high-mass star. It has been proposed that the luminosity of 6.7~GHz methanol masers increases as the associated young stellar object evolves \citep[e.g.][]{Breen+10a}. In this scenario sources such as G\,339.884$-$1.259, which is amongst the most luminous 6.7~GHz methanol masers in the galaxy and is associated with a number of rare class~II methanol transitions \citep{Krishnan+13,Ellingsen+13}, are close to the end of the methanol-maser phase of high-mass star formation. Recently, \citet{Urquhart+15} have put forward an alternative interpretation, that the methanol maser luminosity has a closer dependence on the bolometric luminosity of the associated high-mass star than its evolutionary phase. In this case, we would expect G\,339.884$-$1.259 to be associated with a high-mass young stellar object towards the upper end of the mass-range. However, this is not the case and demonstrates that high luminosity, multiple transition methanol maser emission need not be associated with the most massive O-type young stars.  A single example is insufficient to resolve the issue of whether evolutionary stage or stellar mass plays the primary role in determining class~II methanol maser luminosity, however, G\,339.884$-$1.259 shows greater consistency with the expectations of the \citet{Breen+10a} evolutionary hypothesis.

\begin{table}
\centering
\caption{Physical parameters of G\,339.884$-$1.259 as described in Section~\ref{sec:pcis} and adjusted to a preferred distance of 2.1 kpc.}
\begin{tabular}{ccccc}\hline
  $n_e$ &  $M_{\mbox{\mbox{\tiny \ionhy}}} $& $U$ & $\log ~N_L$ & Spectral\\
 (cm$^{-3}$) & (\msol )       & (pc~cm$^{-2}$) &  (s$^{-1}$)   & type \\
\hline
\hline
$3.1 \times 10^4 $ & $6 \times 10^{-4}$ & 5.7 & 45.6         & B1 \\
\hline
 \end{tabular}
  \label{tab:starType}
\end{table}

\section{Conclusion}
We are currently conducting a large project using the LBA to measure the positions of thirty, 6.7~GHz methanol masers relative to background quasars, in the Southern hemisphere to sub-milliarcsecond accuracy. These measurements will be used to determine their distances using trigonometric parallax. The source list contains many prominent HMSFRs in the third and fourth quadrants of the Milky Way galaxy, where the LBA's performance is unmatched in its astrometric capabilities as a VLBI instrument. In this paper, we have shown the potential of this project by successfully making the first parallax measurements to a southern 6.7~GHz methanol maser source. The parallax of 0.48$\pm $0.08~mas to G\,339.884$-$1.259 corresponds to a distance of $2.1^{+0.4}_{-0.3}$ kpc. In combining this result with measurements of other HMSFRs in \citet{Sato+14}, we place G\,339.884$-$1.259 at the near edge of the Scutum spiral arm of the Milky Way, and determine an updated pitch angle of $\psi = $ 19.2$^\circ \pm $4.1$^\circ $ for this arm.

We have used the parallax distance to update the estimated physical parameters for the G339.884$-$1.259 star formation region and now classify it to be of spectral type B1. The young stellar object associated with G\,339.884$-$1.259 is relatively small for a ``high-mass'' object and is a strong example of the uncorrelated nature between stellar mass and intensity of 6.7~GHz maser emission.

Results from VLBI astrometric observations including the BeSSeL survey are continuing to shape the way we view our Galaxy, by clearly revealing its spiral arm structure, rotation dynamics and mass. These results are based on observations which are concentrated in the first and second quadrants of the Milky Way \citep{Reid+14a}, and the inclusion of results from the LBA is vital to ensure that sources are sampled from across a broad range of Galactic longitudes in order for a complete picture of the Galaxy to emerge.

\section*{Acknowledgements}
This paper is dedicated to the memory of our co-author James Caswell who passed away January 14$^{\mbox{\scriptsize th}}$ 2015. James was a leading figure in maser astronomy for more than three decades and his work in this field represents a peerless legacy to future researchers.

Funding assistance was provided in part by Sigma Xi Grants-in-Aid of research, the Deutscher Akademischer Austauschdienst (DAAD) and National Natural Science Foundation of China (Grant No. 11133008).

The authors would like to thank the referee for their detailed analysis and comments in reviewing this paper.

\appendix
\section{Tropospheric and Ionospheric Contributions to Astrometric Accuracy}
\label{sec:icrfObs}
We found that the parallax measurement with the smallest error floors was obtained when troposphere corrections were applied to the 2013 March and 2013 August epochs only. When the corrections were incorporated for all epochs, the parallax was measured to be 0.58$\pm $0.11 mas. When we excluded the tropospheric corrections for all epochs the measured trigonometric parallax was 0.50$\pm $0.13 mas. We attribute the effectiveness of applying this calibration in each epoch to limitations that we encountered in estimating both the tropospheric and ionospheric delay.

\subsection{Troposphere Calibration}
The multiband delay solutions obtained from 4 $\times $ 16 MHz IFs were found to be significantly noisier than those obtained when only 2 $\times $ 16 MHz IFs were used. As a result, the multiband delays for our observations were determined from a total bandwidth separation of only 32 MHz instead of the intended 400 MHz. This led to a loss in the accuracy of the delay measurement and affected our ability to model the troposphere path length contribution.

\citet{Deller+09b} demonstrated that the available TEC models, for ionosphere delay correction in the Southern hemisphere, are sometimes in error due to the lower density of GPS stations in this region. The ionospheric delay is dispersive whereas the tropospheric delay is not. Hence uncorrected ionospheric delays which are inadvertently handled as non-dispersive tropospheric delays can result in erroneous phase corrections \citep{Rygl+10}. This could be the reason why the search for multiband delays over the highly sensitive 400 MHz range was noisy, and we detail the ionospheric path length contribution in the next subsection.

In addition to the troposphere path length, the clock drift over the course of the observations was also resolved from the multiband delays. The clock model assumes that the H-maser at each station had a delay drift which varied linearly with time. Figure~\ref{fig:geoBlk} shows the residual differences between the measured clock and modeled offsets for the duration of an observation. The squares, circles and crosses represent (respectively) the data, the model and residuals between the two. The near zero scatter of the residuals indicates that a linear model for the clock drift at each station was successful. The RMS noise of the residuals were found to be between 0.03 to 0.3 nanoseconds across all baselines and epochs.

\renewcommand{\thefigure}{A\arabic{figure}}
\setcounter{figure}{0}
\begin{figure}
  \centering
     \includegraphics[width=\linewidth ]{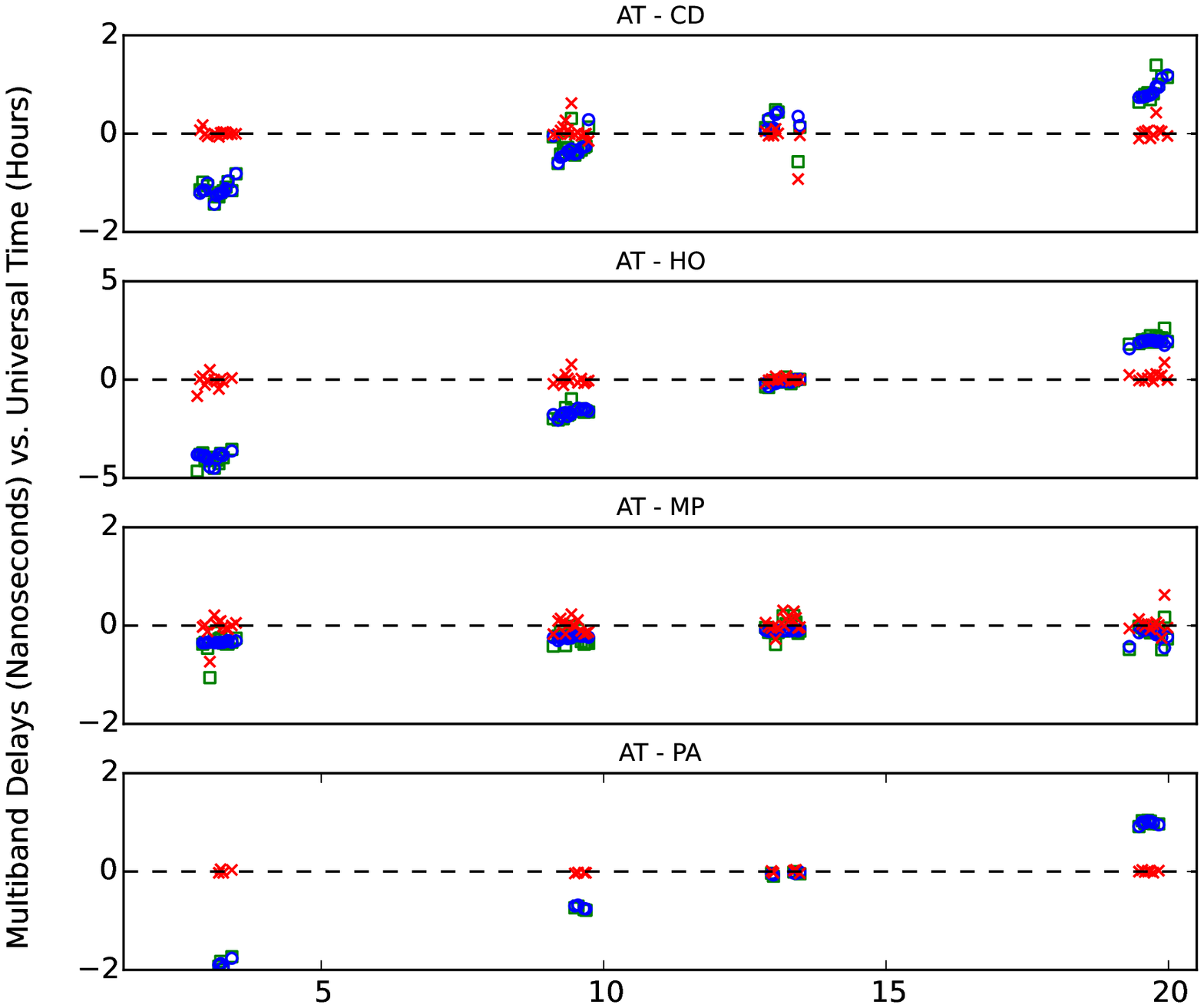}
\caption{A sample of multi-band delay solutions from Ceduna (CD), Hobart (HO), Mopra (MP) and Parkes (PA) to the ATCA (AT), determined from the ICRF mode observations in 2013 June. The squares, circles and crosses represent (respectively) the data, the model and residuals between the two.}
  \label{fig:geoBlk}
\end{figure}

The hetrogeneous nature of the array means that some antennas were able to participate in a relatively small number of ICRF scans, Parkes due to a combination of slow slew rates, limited elevation coverage and HartRAO due to its distance from the rest of the LBA antennas. We were not able to include HartRAO in the majority of the ICRF observations as the quasars which had risen at the Australian stations were often set here. As a result, we were unable to determine the clock drift for HartRAO and have not included observations from this station to determine the results given in Table~\ref{tab:parRes}. We are currently investigating alternative methods to determine the clock drift rate from HartRAO for inclusion in the future.

\subsection{Ionosphere calibration}
Given the observing parameters (6.7~GHz, 4 minute cycle time, 2.48$^\circ $ separation between the calibrator and the target, 3 cm residual zenith path length \citep{Reid+99} and residual ionospheric content of 6 TECU \citep{Ho+97}) we can estimate the expected error contributions from static and dynamic components of the troposphere and the ionosphere using the formulae in \citet{Asaki+07}. This predicts a dynamic tropospheric phase error of 28$^\circ $, a static tropospheric phase error of
15$^\circ $, a dynamic ionospheric phase error of 5$^\circ $ and a static ionospheric phase error of 21$^\circ $. The geodetic blocks typically reduce the residual zenith path length to about 1 cm, which would decrease the static tropospheric phase error to 5$^\circ $. Dynamic errors, because of the short timescale on which these operate, reduce the measured flux density of the targeted source without having a large effect on the positional centroid. This clearly leaves the residual static ionosphere as the largest uncorrected effect. Static residuals will introduce a shift in the observed position, as measured on any single baseline. For arrays with 10 or so antennas these shifts average out somewhat, but the LBA with 5 to 6 antennas is more vulnerable to this effect.

Tuning the observational parameters, such as the cycle time and the separation between sources, is the most effective method to reduce these error contributions. For example, reducing the cycle time to 60 seconds would diminish the dynamic phase errors to less than 10$^\circ $ and decreasing the separation between the calibrator and the target to 1$^\circ $ would cut back the static phase errors to less than
10$^\circ $. However, the array sensitivity places limits the minimum useful scan duration and the separation of the closest suitable phase calibrator. In this case we were not able to have shorter scans or closer calibrators.

For observations at higher frequencies, such as 22~GHz, the contribution of the residual zenith path length dominates. Therefore the geodetic blocks are absolutely essential for the measurements in \citet{Reid+14a,Honma+12} as they reduce the typical errors from 3 cm to about 1 cm. However at 6.7~GHz the residual ionospheric contribution of 6 TECU is equivalent to 5.3 cm, as opposed to 0.5 cm at 22~GHz, and there is currently no established strategy for minimizing these contributions. Alternative approaches to lower the residual ionospheric contribution are currently being investigated and will be the subject of future publications.

\section{Delay Calibration for Phase Referencing Data}
\label{sec:delayCal}
We used FRING to determine the visibility phases and group delay from an individual scan on the delay calibrator J\,1706$-$4600 taken from the continuum mode data. It is assumed that the delay $\tau $ calculated by FRING is a constant for the IF and can be used to determine a phase correction $\Delta \phi $ for a frequency channel with width $\Delta \nu $ and with respect to the lower-band edge given by

\begin{align}
\Delta \phi ~=~ \tau \Delta \nu \nonumber
\end{align}

The 2~MHz maser zoom-band is a sub-band of the IF used to determine the manual delay, but it has a different lower band-edge (offset by $B_{off} $) to the continuum data and a different spectral channel width $\delta \nu $.  To apply the appropriate phase correction to the maser reference channel it was necessary to edit the FQ table in AIPS so that the spectral channel width $\delta \nu^{\prime}$ times the channel number $\sigma _{pr}$ corresponds to the frequency of the maser phase reference channel $\nu _ {pr} $, referenced to the lower band-edge of the continuum mode data given by

\begin{align*}
\nu _{pr} ~&=~ B_{off} + \sigma _{pr} \delta \nu =~ \sigma _{pr} \delta \nu ^\prime \\ \nonumber
\tf  \delta \nu ^\prime ~&=~ \frac{B_{off}}{\sigma _{pr}} + \delta \nu \\ \nonumber
\end{align*}

It was also necessary to edit the total bandwidth parameter in the FQ table for the maser data so that it matched the bandwidth of the continuum mode data.

This step can be avoided if the delay calibrator scans are correlated for both continuum and zoom mode configurations and it is recommended that this be the general procedure in future LBA observations. Equivalent solutions for $\Delta \phi $ and $\tau $ can be obtained for both datasets if an identical time range is used in FRING. An alternative approach would be to correlate the continuum data into several channels such that one of these corresponds exactly to the band coverage of the zoom mode data. The multi-band group delay solutions obtained from the continuum mode data could then be used to find the phase solution in the channel corresponding to the zoom band. This solution could then be directly transferred to the zoom band data.

\bibliography{para.bib}

\begin{thebibliography}{}
\expandafter\ifx\csname natexlab\endcsname\relax\def\natexlab#1{#1}\fi

\bibitem[{{Alef}(1988)}]{Alef+88}
{Alef}, W. 1988, in \iaus, Vol. 129, The Impact of VLBI on Astrophysics and
  Geophysics, ed. M.~J. {Reid} \& J.~M. {Moran}, 523

\bibitem[{{Asaki} {et~al.}(2007){Asaki}, {Sudou}, {Kono}, {Doi}, {Dodson},
  {Pradel}, {Murata}, {Mochizuki}, {Edwards}, {Sasao}, \&
  {Fomalont}}]{Asaki+07}
{Asaki}, Y., {Sudou}, H., {Kono}, Y., {et~al.} 2007, \pasj, 59, 397

\bibitem[{{Beasley} \& {Conway}(1995)}]{Beasley+95}
{Beasley}, A.~J., \& {Conway}, J.~E. 1995, in \aspc, Vol.~82, Very Long
  Baseline Interferometry and the VLBA, ed. J.~A. {Zensus}, P.~J. {Diamond}, \&
  P.~J. {Napier}, 327

\bibitem[{{Bonnell} {et~al.}(1998){Bonnell}, {Bate}, \&
  {Zinnecker}}]{Bonnell+98}
{Bonnell}, I.~A., {Bate}, M.~R., \& {Zinnecker}, H. 1998, \mnras, 298, 93

\bibitem[{{Breen} {et~al.}(2010){Breen}, {Ellingsen}, {Caswell}, \&
  {Lewis}}]{Breen+10a}
{Breen}, S.~L., {Ellingsen}, S.~P., {Caswell}, J.~L., \& {Lewis}, B.~E. 2010,
  \mnras, 401, 2219

\bibitem[{{Bronfman} {et~al.}(1996){Bronfman}, {Nyman}, \& {May}}]{Bronfman+96}
{Bronfman}, L., {Nyman}, L.-A., \& {May}, J. 1996, \aaps, 115, 81

\bibitem[{{Brunthaler} {et~al.}(2009){Brunthaler}, {Reid}, {Menten}, {Zheng},
  {Moscadelli}, \& {Xu}}]{Brunthaler+09}
{Brunthaler}, A., {Reid}, M.~J., {Menten}, K.~M., {et~al.} 2009, \apj, 693, 424

\bibitem[{{Brunthaler} {et~al.}(2011){Brunthaler}, {Reid}, {Menten}, {Zheng},
  {Bartkiewicz}, {Choi}, {Dame}, {Hachisuka}, {Immer}, {Moellenbrock},
  {Moscadelli}, {Rygl}, {Sanna}, {Sato}, {Wu}, {Xu}, \&
  {Zhang}}]{Brunthaler+11}
---. 2011, \an, 332, 461

\bibitem[{{Caswell} \& {Haynes}(1983)}]{Caswell+83}
{Caswell}, J.~L., \& {Haynes}, R.~F. 1983, \aujph, 36, 361

\bibitem[{{Caswell} {et~al.}(1975){Caswell}, {Murray}, {Roger}, {Cole}, \&
  {Cooke}}]{Caswell+75}
{Caswell}, J.~L., {Murray}, J.~D., {Roger}, R.~S., {Cole}, D.~J., \& {Cooke},
  D.~J. 1975, \aap, 45, 239

\bibitem[{{Caswell} {et~al.}(2000){Caswell}, {Yi}, {Booth}, \&
  {Cragg}}]{Caswell+00}
{Caswell}, J.~L., {Yi}, J., {Booth}, R.~S., \& {Cragg}, D.~M. 2000, \mnras,
  313, 599

\bibitem[{{Caswell} {et~al.}(2010){Caswell}, {Fuller}, {Green}, {Avison},
  {Breen}, {Brooks}, {Burton}, {Chrysostomou}, {Cox}, {Diamond}, {Ellingsen},
  {Gray}, {Hoare}, {Masheder}, {McClure-Griffiths}, {Pestalozzi}, {Phillips},
  {Quinn}, {Thompson}, {Voronkov}, {Walsh}, {Ward-Thompson}, {Wong-McSweeney},
  {Yates}, \& {Cohen}}]{Caswell+10}
{Caswell}, J.~L., {Fuller}, G.~A., {Green}, J.~A., {et~al.} 2010, \mnras, 404,
  1029

\bibitem[{{Caswell} {et~al.}(2011){Caswell}, {Fuller}, {Green}, {Avison},
  {Breen}, {Ellingsen}, {Gray}, {Pestalozzi}, {Quinn}, {Thompson}, \&
  {Voronkov}}]{Caswell+11}
---. 2011, \mnras, 417, 1964

\bibitem[{{Chibueze} {et~al.}(2014){Chibueze}, {Omodaka}, {Handa}, {Imai},
  {Kurayama}, {Nagayama}, {Sunada}, {Nakano}, {Hirota}, \&
  {Honma}}]{Chibueze+14}
{Chibueze}, J.~O., {Omodaka}, T., {Handa}, T., {et~al.} 2014, \apj, 784, 114

\bibitem[{{Choi} {et~al.}(2014){Choi}, {Hachisuka}, {Reid}, {Xu}, {Brunthaler},
  {Menten}, \& {Dame}}]{Choi+14}
{Choi}, Y.~K., {Hachisuka}, K., {Reid}, M.~J., {et~al.} 2014, \apj, 790, 99

\bibitem[{{Cohen} {et~al.}(1980){Cohen}, {Cong}, {Dame}, \&
  {Thaddeus}}]{Cohen+80}
{Cohen}, R.~S., {Cong}, H., {Dame}, T.~M., \& {Thaddeus}, P. 1980, \apjl, 239,
  L53

\bibitem[{{Cragg} {et~al.}(1992){Cragg}, {Johns}, {Godfrey}, \&
  {Brown}}]{Cragg+92}
{Cragg}, D.~M., {Johns}, K.~P., {Godfrey}, P.~D., \& {Brown}, R.~D. 1992,
  \mnras, 259, 203

\bibitem[{{Dame} {et~al.}(2001){Dame}, {Hartmann}, \& {Thaddeus}}]{Dame+01}
{Dame}, T.~M., {Hartmann}, D., \& {Thaddeus}, P. 2001, \apj, 547, 792

\bibitem[{{De Buizer} {et~al.}(2002){De Buizer}, {Walsh}, {Pi{\~n}a},
  {Phillips}, \& {Telesco}}]{deBuizer+02}
{De Buizer}, J.~M., {Walsh}, A.~J., {Pi{\~n}a}, R.~K., {Phillips}, C.~J., \&
  {Telesco}, C.~M. 2002, \apj, 564, 327

\bibitem[{{Deller} {et~al.}(2009{\natexlab{a}}){Deller}, {Tingay}, {Bailes}, \&
  {Reynolds}}]{Deller+09b}
{Deller}, A.~T., {Tingay}, S.~J., {Bailes}, M., \& {Reynolds}, J.~E.
  2009{\natexlab{a}}, \apj, 701, 1243

\bibitem[{{Deller} {et~al.}(2009{\natexlab{b}}){Deller}, {Tingay}, \&
  {Brisken}}]{Deller+09a}
{Deller}, A.~T., {Tingay}, S.~J., \& {Brisken}, W. 2009{\natexlab{b}}, \apj,
  690, 198

\bibitem[{{Deller} {et~al.}(2011){Deller}, {Brisken}, {Phillips}, {Morgan},
  {Alef}, {Cappallo}, {Middelberg}, {Romney}, {Rottmann}, {Tingay}, \&
  {Wayth}}]{Deller+11}
{Deller}, A.~T., {Brisken}, W.~F., {Phillips}, C.~J., {et~al.} 2011, \pasp,
  123, 275

\bibitem[{{Dodson}(2008)}]{Dodson+08}
{Dodson}, R. 2008, \aap, 480, 767

\bibitem[{{Dodson} {et~al.}(2003){Dodson}, {Legge}, {Reynolds}, \&
  {McCulloch}}]{Dodson+03}
{Dodson}, R., {Legge}, D., {Reynolds}, J.~E., \& {McCulloch}, P.~M. 2003, \apj,
  596, 1137

\bibitem[{{Dodson} {et~al.}(2014){Dodson}, {Rioja}, {Jung}, {Sohn}, {Byun},
  {Cho}, {Lee}, {Kim}, {Kim}, {Oh}, {Han}, {Je}, {Chung}, {Wi}, {Kang}, {Lee},
  {Chung}, {Kim}, {Kim}, {Lee}, {Roh}, {Oh}, {Yeom}, {Song}, \&
  {Kang}}]{Dodson+14}
{Dodson}, R., {Rioja}, M.~J., {Jung}, T.-H., {et~al.} 2014, \aj, 148, 97

\bibitem[{{Ellingsen} {et~al.}(2011){Ellingsen}, {Breen}, {Sobolev},
  {Voronkov}, {Caswell}, \& {Lo}}]{Ellingsen+11a}
{Ellingsen}, S.~P., {Breen}, S.~L., {Sobolev}, A.~M., {et~al.} 2011, \apj, 742,
  109

\bibitem[{{Ellingsen} {et~al.}(2013){Ellingsen}, {Breen}, {Voronkov}, \&
  {Dawson}}]{Ellingsen+13}
{Ellingsen}, S.~P., {Breen}, S.~L., {Voronkov}, M.~A., \& {Dawson}, J.~R. 2013,
  \mnras, 429, 3501

\bibitem[{{Ellingsen} {et~al.}(2004){Ellingsen}, {Cragg}, {Lovell}, {Sobolev},
  {Ramsdale}, \& {Godfrey}}]{Ellingsen+04}
{Ellingsen}, S.~P., {Cragg}, D.~M., {Lovell}, J.~E.~J., {et~al.} 2004, \mnras,
  354, 401

\bibitem[{{Ellingsen} {et~al.}(1996){Ellingsen}, {Norris}, \&
  {McCulloch}}]{Ellingsen+96}
{Ellingsen}, S.~P., {Norris}, R.~P., \& {McCulloch}, P.~M. 1996, \mnras, 279,
  101

\bibitem[{{Ellingsen} {et~al.}(2005){Ellingsen}, {Shabala}, \&
  {Kurtz}}]{Ellingsen+05}
{Ellingsen}, S.~P., {Shabala}, S.~S., \& {Kurtz}, S.~E. 2005, \mnras, 357, 1003

\bibitem[{{Fomalont}(2013)}]{Fomalont+13}
{Fomalont}, E.~B. 2013, Astrometry for Astrophysics: Methods, Models, and
  Applications (Cambridge University Press), 175--198

\bibitem[{{Garay} \& {Lizano}(1999)}]{Garay+99}
{Garay}, G., \& {Lizano}, S. 1999, \pasp, 111, 1049

\bibitem[{{Goddi} {et~al.}(2011){Goddi}, {Moscadelli}, \& {Sanna}}]{Goddi+11}
{Goddi}, C., {Moscadelli}, L., \& {Sanna}, A. 2011, \aap, 535, L8

\bibitem[{{Green} \& {McClure-Griffiths}(2011)}]{Green+11}
{Green}, J.~A., \& {McClure-Griffiths}, N.~M. 2011, \mnras, 417, 2500

\bibitem[{{Green} {et~al.}(2010){Green}, {Caswell}, {Fuller}, {Avison},
  {Breen}, {Ellingsen}, {Gray}, {Pestalozzi}, {Quinn}, {Thompson}, \&
  {Voronkov}}]{Green+10}
{Green}, J.~A., {Caswell}, J.~L., {Fuller}, G.~A., {et~al.} 2010, \mnras, 409,
  913

\bibitem[{{Green} {et~al.}(2012){Green}, {Caswell}, {Fuller}, {Avison},
  {Breen}, {Ellingsen}, {Gray}, {Pestalozzi}, {Quinn}, {Thompson}, \&
  {Voronkov}}]{Green+12a}
---. 2012, \mnras, 420, 3108

\bibitem[{{Greisen}(2003)}]{Greisen+03}
{Greisen}, E.~W. 2003, \iha, 285, 109

\bibitem[{{Hachisuka} {et~al.}(2006){Hachisuka}, {Brunthaler}, {Menten},
  {Reid}, {Imai}, {Hagiwara}, {Miyoshi}, {Horiuchi}, \& {Sasao}}]{Hachisuka+06}
{Hachisuka}, K., {Brunthaler}, A., {Menten}, K.~M., {et~al.} 2006, \apj, 645,
  337

\bibitem[{{Hancock} {et~al.}(2011){Hancock}, {Roberts}, {Kesteven}, {Ekers},
  {Sadler}, {Murphy}, {Massardi}, {Ricci}, {Calabretta}, {de Zotti}, {Edwards},
  {Ekers}, {Jackson}, {Leach}, {Phillips}, {Sault}, {Staveley-Smith},
  {Subrahmanyan}, {Walker}, \& {Wilson}}]{Hancock+11}
{Hancock}, P.~J., {Roberts}, P., {Kesteven}, M.~J., {et~al.} 2011, \exa, 32,
  147

\bibitem[{{Ho} {et~al.}(1997){Ho}, {Wilson}, {Mannucci}, {Lindqwister}, \&
  {Yuan}}]{Ho+97}
{Ho}, C.~M., {Wilson}, B.~D., {Mannucci}, A.~J., {Lindqwister}, U.~J., \&
  {Yuan}, D.~N. 1997, \rasc, 32, 1499

\bibitem[{{Honma} {et~al.}(2008){Honma}, {Tamura}, \& {Reid}}]{Honma+08}
{Honma}, M., {Tamura}, Y., \& {Reid}, M.~J. 2008, \pasj, 60, 951

\bibitem[{{Honma} {et~al.}(2004){Honma}, {Bushimata}, {Choi}, {Fujii},
  {Hirota}, {Horiai}, {Imai}, {Inomata}, {Ishitsuka}, {Iwadate}, {Jike},
  {Kameya}, {Kamohara}, {Kan-Ya}, {Kawaguchi}, {Kobayashi}, {Kuji}, {Kurayama},
  {Manabe}, {Miyaji}, {Nakagawa}, {Nakashima}, {Omodaka}, {Oyama}, {Sakai},
  {Sato}, {Sasao}, {Shibata}, {Shimizu}, {Sora}, {Suda}, {Tamura}, \&
  {Yamashita}}]{Honma+04}
{Honma}, M., {Bushimata}, T., {Choi}, Y.~K., {et~al.} 2004, in \evnconf, ed.
  R.~{Bachiller}, F.~{Colomer}, J.-F. {Desmurs}, \& P.~{de Vicente}, 203--204

\bibitem[{{Honma} {et~al.}(2007){Honma}, {Bushimata}, {Choi}, {Hirota}, {Imai},
  {Iwadate}, {Jike}, {Kameya}, {Kamohara}, {Kan-Ya}, {Kawaguchi}, {Kijima},
  {Kobayashi}, {Kuji}, {Kurayama}, {Manabe}, {Miyaji}, {Nagayama}, {Nakagawa},
  {Oh}, {Omodaka}, {Oyama}, {Sakai}, {Sato}, {Sasao}, {Shibata}, {Shintani},
  {Suda}, {Tamura}, {Tsushima}, \& {Yamashita Kazuyoshi}}]{Honma+07}
{Honma}, M., {Bushimata}, T., {Choi}, Y.~K., {et~al.} 2007, \pasj, 59, 889

\bibitem[{{Honma} {et~al.}(2012){Honma}, {Nagayama}, {Ando}, {Bushimata},
  {Choi}, {Handa}, {Hirota}, {Imai}, {Jike}, {Kim}, {Kameya}, {Kawaguchi},
  {Kobayashi}, {Kurayama}, {Kuji}, {Matsumoto}, {Manabe}, {Miyaji}, {Motogi},
  {Nakagawa}, {Nakanishi}, {Niinuma}, {Oh}, {Omodaka}, {Oyama}, {Sakai},
  {Sato}, {Sato}, {Shibata}, {Shiozaki}, {Sunada}, {Tamura}, {Ueno}, \&
  {Yamauchi}}]{Honma+12}
{Honma}, M., {Nagayama}, T., {Ando}, K., {et~al.} 2012, \pasj, 64, 136

\bibitem[{{Imai} {et~al.}(2013){Imai}, {Kurayama}, {Honma}, \&
  {Miyaji}}]{Imai+13}
{Imai}, H., {Kurayama}, T., {Honma}, M., \& {Miyaji}, T. 2013, \pasj, 65, 28

\bibitem[{{Jones} \& {Dickey}(2012)}]{Jones+12}
{Jones}, C., \& {Dickey}, J.~M. 2012, \apj, 753, 62

\bibitem[{{Jones} {et~al.}(2013){Jones}, {Dickey}, {Dawson},
  {McClure-Griffiths}, {Anderson}, \& {Bania}}]{Jones+13}
{Jones}, C., {Dickey}, J.~M., {Dawson}, J.~R., {et~al.} 2013, \apj, 774, 117

\bibitem[{{Krishnan} {et~al.}(2013){Krishnan}, {Ellingsen}, {Voronkov}, \&
  {Breen}}]{Krishnan+13}
{Krishnan}, V., {Ellingsen}, S.~P., {Voronkov}, M.~A., \& {Breen}, S.~L. 2013,
  \mnras, 433, 3346

\bibitem[{{Ma} {et~al.}(2009){Ma}, {Arias}, {Bianco}, {Boboltz}, {Bolotin},
  {Charlot}, {Engelhardt}, {Fey}, A., {Gontier}, {Heinkelmann}, {Jacobs},
  {Kurdubov}, {Lambert}, {Malkin}, {Nothnagel}, {Petrov}, {Skurikhina},
  {Sokolova}, {Souchay}, {Sovers}, {Tesmer}, {Titov}, {Wang}, {Zharov},
  {Barache}, {B\"{o}ckmann}, {Collioud}, {Gipson}, {Gordon}, {Lytvyn},
  {MacMillan}, {Ojha}, {Fey}, {Gordon}, {Jacobs}, {Bianco}, {Boboltz},
  {Bolotin}, {Charlot}, {Engelhardt}, {Fey}, A., {Gontier}, {Heinkelmann},
  {Jacobs}, {Kurdubov}, {Lambert}, {Malkin}, {Nothnagel}, {Petrov},
  {Skurikhina}, {Sokolova}, {Souchay}, {Sovers}, {Tesmer}, {Titov}, {Wang},
  {Zharov}, {Barache}, {B\"{o}ckmann}, {Collioud}, {Gipson}, {Gordon},
  {Lytvyn}, {MacMillan}, {Ojha}, {Fey}, {Gordon}, \& {Jacobs}}]{Ma+09}
{Ma}, C., {Arias}, E.~F., {Bianco}, G., {et~al.} 2009, The Second Realization
  of the International Celestial Reference Frame by Very Long Baseline
  Interferometry, \itn~35, International Earth Rotation and Reference System
  Service (IERS), International VLBI Service for Geodesy and Astrometry (IVS)

\bibitem[{{McConnell} {et~al.}(2012){McConnell}, {Sadler}, {Murphy}, \&
  {Ekers}}]{McConnell+12}
{McConnell}, D., {Sadler}, E.~M., {Murphy}, T., \& {Ekers}, R.~D. 2012, \mnras,
  422, 1527

\bibitem[{{McKee} \& {Tan}(2003)}]{McKee+03}
{McKee}, C.~F., \& {Tan}, J.~C. 2003, \apj, 585, 850

\bibitem[{{Moscadelli} \& {Goddi}(2014)}]{Moscadelli+14}
{Moscadelli}, L., \& {Goddi}, C. 2014, \aap, 566, A150

\bibitem[{{Moscadelli} {et~al.}(2009){Moscadelli}, {Reid}, {Menten},
  {Brunthaler}, {Zheng}, \& {Xu}}]{Moscadelli+09}
{Moscadelli}, L., {Reid}, M.~J., {Menten}, K.~M., {et~al.} 2009, \apj, 693, 406

\bibitem[{{Norris} {et~al.}(1987){Norris}, {Caswell}, {Gardner}, \&
  {Wellington}}]{Norris+87}
{Norris}, R.~P., {Caswell}, J.~L., {Gardner}, F.~F., \& {Wellington}, K.~J.
  1987, \apjl, 321, L159

\bibitem[{{Norris} {et~al.}(1993){Norris}, {Whiteoak}, {Caswell}, {Wieringa},
  \& {Gough}}]{Norris+93}
{Norris}, R.~P., {Whiteoak}, J.~B., {Caswell}, J.~L., {Wieringa}, M.~H., \&
  {Gough}, R.~G. 1993, \apj, 412, 222

\bibitem[{{Norris} {et~al.}(1998){Norris}, {Byleveld}, {Diamond}, {Ellingsen},
  {Ferris}, {Gough}, {Kesteven}, {McCulloch}, {Phillips}, {Reynolds},
  {Tzioumis}, {Takahashi}, {Troup}, \& {Wellington}}]{Norris+98}
{Norris}, R.~P., {Byleveld}, S.~E., {Diamond}, P.~J., {et~al.} 1998, \apj, 508,
  275

\bibitem[{{Panagia}(1973)}]{Panagia+73}
{Panagia}, N. 1973, \aj, 78, 929

\bibitem[{{Panagia} \& {Walmsley}(1978)}]{Panagia+78}
{Panagia}, N., \& {Walmsley}, C.~M. 1978, \aap, 70, 411

\bibitem[{{Petrov} {et~al.}(2011){Petrov}, {Phillips}, {Bertarini}, {Murphy},
  \& {Sadler}}]{Petrov+11}
{Petrov}, L., {Phillips}, C., {Bertarini}, A., {Murphy}, T., \& {Sadler}, E.~M.
  2011, \mnras, 414, 2528

\bibitem[{{Phillips} {et~al.}(1998){Phillips}, {Norris}, {Ellingsen}, \&
  {McCulloch}}]{Phillips+98}
{Phillips}, C.~J., {Norris}, R.~P., {Ellingsen}, S.~P., \& {McCulloch}, P.~M.
  1998, \mnras, 300, 1131

\bibitem[{{Reid} \& {Honma}(2014)}]{Reid+14b}
{Reid}, M.~J., \& {Honma}, M. 2014, \araa, 52, 339

\bibitem[{{Reid} {et~al.}(2009{\natexlab{a}}){Reid}, {Menten}, {Brunthaler},
  {Zheng}, {Moscadelli}, \& {Xu}}]{Reid+09a}
{Reid}, M.~J., {Menten}, K.~M., {Brunthaler}, A., {et~al.} 2009{\natexlab{a}},
  \apj, 693, 397

\bibitem[{{Reid} {et~al.}(1999){Reid}, {Readhead}, {Vermeulen}, \&
  {Treuhaft}}]{Reid+99}
{Reid}, M.~J., {Readhead}, A.~C.~S., {Vermeulen}, R.~C., \& {Treuhaft}, R.~N.
  1999, \apj, 524, 816

\bibitem[{{Reid} {et~al.}(2009{\natexlab{b}}){Reid}, {Menten}, {Zheng},
  {Brunthaler}, {Moscadelli}, {Xu}, {Zhang}, {Sato}, {Honma}, {Hirota},
  {Hachisuka}, {Choi}, {Moellenbrock}, \& {Bartkiewicz}}]{Reid+09b}
{Reid}, M.~J., {Menten}, K.~M., {Zheng}, X.~W., {et~al.} 2009{\natexlab{b}},
  \apj, 700, 137

\bibitem[{{Reid} {et~al.}(2014){Reid}, {Menten}, {Brunthaler}, {Zheng}, {Dame},
  {Xu}, {Wu}, {Zhang}, {Sanna}, {Sato}, {Hachisuka}, {Choi}, {Immer},
  {Moscadelli}, {Rygl}, \& {Bartkiewicz}}]{Reid+14a}
{Reid}, M.~J., {Menten}, K.~M., {Brunthaler}, A., {et~al.} 2014, \apj, 783, 130

\bibitem[{{Rioja} {et~al.}(2008){Rioja}, {Dodson}, {Kamohara}, {Colomer},
  {Bujarrabal}, \& {Kobayashi}}]{Rioja+08}
{Rioja}, M.~J., {Dodson}, R., {Kamohara}, R., {et~al.} 2008, \pasj, 60, 1031

\bibitem[{{Rygl} {et~al.}(2008){Rygl}, {Brunthaler}, {Menten}, {Reid}, \& {van
  Langevelde}}]{Rygl+08}
{Rygl}, K.~L.~J., {Brunthaler}, A., {Menten}, K.~M., {Reid}, M.~J., \& {van
  Langevelde}, H.~J. 2008, in \evncon

\bibitem[{{Rygl} {et~al.}(2010){Rygl}, {Brunthaler}, {Reid}, {Menten}, {van
  Langevelde}, \& {Xu}}]{Rygl+10}
{Rygl}, K.~L.~J., {Brunthaler}, A., {Reid}, M.~J., {et~al.} 2010, \aap, 511, A2

\bibitem[{{Sakai} {et~al.}(2012){Sakai}, {Honma}, {Nakanishi}, {Sakanoue},
  {Kurayama}, {Shibata}, \& {Shizugami}}]{Sakai+12}
{Sakai}, N., {Honma}, M., {Nakanishi}, H., {et~al.} 2012, \pasj, 64, 108

\bibitem[{{Sanna} {et~al.}(2009){Sanna}, {Reid}, {Moscadelli}, {Dame},
  {Menten}, {Brunthaler}, {Zheng}, \& {Xu}}]{Sanna+09}
{Sanna}, A., {Reid}, M.~J., {Moscadelli}, L., {et~al.} 2009, \apj, 706, 464

\bibitem[{{Sanna} {et~al.}(2014){Sanna}, {Reid}, {Menten}, {Dame}, {Zhang},
  {Sato}, {Brunthaler}, {Moscadelli}, \& {Immer}}]{Sanna+14}
{Sanna}, A., {Reid}, M.~J., {Menten}, K.~M., {et~al.} 2014, \apj, 781, 108

\bibitem[{{Sato} {et~al.}(2014){Sato}, {Wu}, {Immer}, {Zhang}, {Sanna}, {Reid},
  {Dame}, {Brunthaler}, \& {Menten}}]{Sato+14}
{Sato}, M., {Wu}, Y.~W., {Immer}, K., {et~al.} 2014, \apj, 793, 72

\bibitem[{{Sewilo} {et~al.}(2004){Sewilo}, {Watson}, {Araya}, {Churchwell},
  {Hofner}, \& {Kurtz}}]{Sewilo+04}
{Sewilo}, M., {Watson}, C., {Araya}, E., {et~al.} 2004, \apjs, 154, 553

\bibitem[{{Sobolev} {et~al.}(1997){Sobolev}, {Cragg}, \&
  {Godfrey}}]{Sobolev+97}
{Sobolev}, A.~M., {Cragg}, D.~M., \& {Godfrey}, P.~D. 1997, \aap, 324, 211

\bibitem[{{Sugiyama} {et~al.}(2014){Sugiyama}, {Fujisawa}, {Doi}, {Honma},
  {Kobayashi}, {Murata}, {Motogi}, {Niinuma}, {Ogawa}, {Wajima},
  {Sawada-Satoh}, \& {Ellingsen}}]{Sugiyama+14}
{Sugiyama}, K., {Fujisawa}, K., {Doi}, A., {et~al.} 2014, \aap, 562, A82

\bibitem[{{Urquhart} {et~al.}(2015){Urquhart}, {Moore}, {Menten}, {K{\"o}nig},
  {Wyrowski}, {Thompson}, {Csengeri}, {Leurini}, \& {Eden}}]{Urquhart+15}
{Urquhart}, J.~S., {Moore}, T.~J.~T., {Menten}, K.~M., {et~al.} 2015, \mnras,
  446, 3461

\bibitem[{{Walker} \& {Chatterjee}(1999)}]{Walker+99}
{Walker}, C., \& {Chatterjee}, S. 1999, {Ionospheric Corrections Using GPS
  Based Models}, VLBA Scientific Memo~23, National Radio Astronomy Observatory
  and Cornell University

\bibitem[{{Walsh} {et~al.}(1998){Walsh}, {Burton}, {Hyland}, \&
  {Robinson}}]{Walsh+98}
{Walsh}, A.~J., {Burton}, M.~G., {Hyland}, A.~R., \& {Robinson}, G. 1998,
  \mnras, 301, 640

\bibitem[{{Westerhout}(1957)}]{Westerhout+57}
{Westerhout}, G. 1957, \ban, 13, 201

\bibitem[{{Wu} {et~al.}(2014){Wu}, {Sato}, {Reid}, {Moscadelli}, {Zhang}, {Xu},
  {Brunthaler}, {Menten}, {Dame}, \& {Zheng}}]{Wu+14}
{Wu}, Y.~W., {Sato}, M., {Reid}, M.~J., {et~al.} 2014, \aap, 566, A17

\bibitem[{{Xu} {et~al.}(2009){Xu}, {Reid}, {Menten}, {Brunthaler}, {Zheng}, \&
  {Moscadelli}}]{Xu+09}
{Xu}, Y., {Reid}, M.~J., {Menten}, K.~M., {et~al.} 2009, \apj, 693, 413

\bibitem[{{Xu} {et~al.}(2006){Xu}, {Reid}, {Zheng}, \& {Menten}}]{Xu+06}
{Xu}, Y., {Reid}, M.~J., {Zheng}, X.~W., \& {Menten}, K.~M. 2006, \sci, 311, 54

\bibitem[{{Xu} {et~al.}(2013){Xu}, {Li}, {Reid}, {Menten}, {Zheng},
  {Brunthaler}, {Moscadelli}, {Dame}, \& {Zhang}}]{Xu+13}
{Xu}, Y., {Li}, J.~J., {Reid}, M.~J., {et~al.} 2013, \apj, 769, 15

\bibitem[{{Zhang} {et~al.}(2013){Zhang}, {Reid}, {Menten}, {Zheng},
  {Brunthaler}, {Dame}, \& {Xu}}]{Zhang+13}
{Zhang}, B., {Reid}, M.~J., {Menten}, K.~M., {et~al.} 2013, \apj, 775, 79

\bibitem[{{Zhang} {et~al.}(2009){Zhang}, {Zheng}, {Reid}, {Menten}, {Xu},
  {Moscadelli}, \& {Brunthaler}}]{Zhang+09}
{Zhang}, B., {Zheng}, X.~W., {Reid}, M.~J., {et~al.} 2009, \apj, 693, 419

\end{thebibliography}
\end{document}